\documentclass[paper,11pt]{JHEP3}
\usepackage{lscape}                                           %
\usepackage[centertags]{amsmath}
\usepackage{amsfonts} \usepackage{amssymb}%\usepackage{eufrak}
\usepackage{colordvi}
\usepackage{graphicx}
\usepackage{lscape}
\newcommand{\COMMENTO}[1]{}
\newcommand{\COMMENTOO}[1]{}
\newcommand{\COMMENTOOK}[1]{}

\newcommand{\cD}{{\cal D}}
\newcommand{\cE}{{\cal E}}

\newcommand{\cG}{{\cal G}}
\newcommand{\cH}{{\cal H}}

\newcommand{\cN}{{\cal N}}
\newcommand{\cR}{{\cal R}}
\newcommand{\cS}{{\cal S}}
\newcommand{\cT}{{\cal T}}

\newcommand{\cV}{{\cal V}}

\newcommand{\uno}{\mathbb{I}}

\newcommand{\R}{\mathbb{R}}

\newcommand{\C}{\mathbb{C}}

\newcommand{\bx}{{\bar x}}
\newcommand{\by}{{\bar y}}

\newcommand{\pp}{{p}}
\newcommand{\bp}{{\bar p}}

\newcommand{\un}[1]{ #1 }

\newcommand{\oh}{\frac{1}{2}}
\newcommand{\Gr}{G}
\title{
The generating function of amplitudes with $N$ twisted and $M$ untwisted states
}

\author{\parbox{11.5cm}{Igor Pesando$^1$}
\\
~\\
~\\
$^1$Dipartimento di Fisica Teorica, Universit\`a di Torino\\
and I.N.F.N. - sezione di Torino \\
Via P. Giuria 1, I-10125 Torino, Italy\\
\vspace{0.3cm}
\email{ipesando@to.infn.it}
}
\abstract{
We show that the generating
function of all amplitudes with $N$ twisted and $M$ untwisted states, i.e. the
Reggeon vertex for magnetized branes on $\R^2$ can be
computed once the correlator of $N$ non excited twisted states
and the corresponding Green function are known and we give an explicit
expression as a functional of the these objects.
}


\preprint{DFTT-20-2011}

\begin{document}

\COMMENTOO{
\begin{enumerate}
\item 
Zero modes?\\
Only one in the chosen gauge $x^2$
\item
How do dipole strings depends on the segment?\\
Moreover $x^i$ are not commutative and $\Theta^{i j}(B)$ depends on
the background. Notice however that $X^{(+)}$ does not depends on $x$
and that $e^{i k  x} |0\rangle = e^{i k  x_0} |0\rangle$ where $x_0$ commute.
\item
Are there two distinguished $\sigma(z)$ and $\bar \sigma (\bar z)$? \\
$Z(e^{i2\pi}z) \rightarrow e^{i2\pi \epsilon} Z(z)$,
$\bar Z(e^{i2\pi}z) \rightarrow e^{-i2\pi \epsilon} \bar Z(z)$,
 implies $ \sigma_\epsilon(x,\bar x)= \sigma_\epsilon(x)
 \sigma_{-\epsilon}(\bar x)$ ?? \\
No, string fields $Z$ and $\bar Z$ depend on the same operators.
\item
Closed strings need $G_{L L}(z,w)$ and $G_{L R}(z, \bar w)$ not only
$G(z, \bar z, w, \bar w)$
\item
Add Stefano and Rofoldo citations
\item
Add B. Nielsson citations
\end{enumerate}
}

\section{Introduction and conclusions}
In the late 80s a lot of work was done in computing the generating
functions of all amplitudes for the bosonic string and superstring.
Many methods were (further) developed such as the sewing method
(\cite{DiVecchia:1986uu}), 
the group theoretic method (\cite{Neveu:1987zf}) 
and conserved charges method (\cite{DiVecchia:1987jb}).
Following the main idea of (\cite{DiVecchia:1986mb},\cite{Petersen:1988cf}) 
in this paper we would like to compute the generating function
for $N$ generic excited twisted states 
and $M$ generic untwisted  states 
on $\R^2$ for the open string 
{ in presence of magnetic fields} in the upper half plane using the path
integral approach.
Much work has been already done in computing non excited twisted
states correlation functions, especially on $T^2$ 
(see for example \cite{Abel:2003yx}, \cite{Cvetic:2003ch},
\cite{Bertolini:2005qh}  and \cite{Duo:2007he}) 
but not so much on the computation of correlators
involving excited twisted fields (\cite{Corrigan:1975sn},
\cite{Hermansson:1990xw} see for earlier work) which remain quite mysterious. 

In this paper we want to show that there is a quite simple way of
labeling excited twisted states which is deeply connected with the
operator-state map and 
that few ingredients are actually needed
for computing all correlators involving excited twisted state and
arbitrary untwisted ones on $\R^2$.
To obtain any correlator is only necessary 
the knowledge of the full (i.e. classical and quantum) 
$N$ non excited twist correlator on the disk\footnote{
The twist fields in this and the following 
correlators are actually $ \sigma_{\epsilon,
  \kappa=0}(x, \bar x)$, see the in the main text.
}
\begin{equation}
C(x_1,\dots x_N)
= 
\langle \sigma_{\epsilon_1}(x_1, \bar x_1)\dots
\sigma_{\epsilon_N}(x_N, \bar x_N)\rangle_{disk,~full}
~~~~x_t\in \R
\end{equation}
and the boundary Green function in presence of such operators
%%%%%%%%%%
\COMMENTOO{
1) Do we only need the boundary Green function? Yes
\\
2) In which conformal gauge is this true?
}
%%%%%%%%%%%
\begin{equation}
\label{Green}
\Gr^{i j}_{bou}(x ; y; \{x_t\}_{t=1\dots N})
=
\Gr^{j i}_{bou}(y ; x; \{x_t\}_{t=1\dots N})
=
\Gr^{i j}(x,\bar x; y, \bar y; \{x_t\}_{t=1\dots N})
\end{equation}
which can be derived from
\begin{equation}
\label{Green-full}
\Gr^{i j}(z,\bar z; w, \bar w; \{x_t\}_{t=1\dots N})
=
\frac{
\langle X^i(z,\bar z) X^j(w, \bar w)
\sigma_{\epsilon_1}(x_1, \bar x_1)\dots \sigma_{\epsilon_N}(x_N, \bar
x_N)\rangle_{disk}
}{
\langle \sigma_{\epsilon_1}(x_1, \bar x_1)\dots
\sigma_{\epsilon_N}(x_N, \bar x_N)\rangle_{disk}
}
.
\end{equation}
by setting $z=x, w=y\in\R$.

The main result of the paper is the generating function for the above
mentioned amplitudes given in eq.s (\ref{Final-Reggeon-N+M}) and
(\ref{Final-Reggeon-N+M-using-cd}).
These two expressions have exactly the same contain but the latter is
written in a more usual way, i.e. using auxiliary expansion variables 
while the former has an expression like those used in the previous
literature (\cite{DiVecchia:1986uu}).
Let us now explain the building blocks of this last version of the  main
formula (\ref{Final-Reggeon-N+M}).

\begin{itemize}
\item
To any (excited) twisted operator inserted at $x_t$
($t=1\dots N$)  in the amplitude we associate an auxiliary Hilbert 
space  $\cH_t$. 
On  $\cH_t$ act the quantum fields $X^i_{(t)}(z,\bar z)$\footnote{
In the following quantum fields have attached the label of the Hilbert
space they act on, e.g. $X^i_{(t)}(z,\bar z)$ while classical fields
in path integral have no label, i.e. $X^i(z,\bar z)$.
} ($i=1,2$ or $i=z,\bar z$) 
%%%%%%%%%%%
\COMMENTOOK{ 
1)  Which is the best way of defining $Z$ as function of  $X^i$?\\
2)  And therefore which metric in space time?\\
Keep oscillators algebra as usual hence $  (d X^1)^2+ (d X^2)^2= 2 d Z d
\bar Z$, i.e. $G_{z \bar z}=1$.}
%%%%%%%%%%%
\begin{align}
\un Z_{(t)}(z,\bar z)
&
= \un X^z_{(t)}(z,\bar z)
= \frac{ \un X^1_{(t)} + i \un X^2_{(t)} }{ \sqrt 2}
= \frac{1}{2}\left( \un Z_{L (t)}(z)+\un Z_{R (t)}(\bar z)\right),
\nonumber\\
\bar{\un Z}_{(t)}(z,\bar z)
&
=  \un X^{\bar z}_{(t)}(z,\bar z)
= \frac { \un X^1_{(t)} - i \un X^2_{(t)}  }{ \sqrt 2}
= \frac{1}{2}\left( \bar{\un Z}_{(t) L}(z)+\bar{\un Z}_{(t) R}(\bar z)\right)
\end{align}
which have expansions
\begin{eqnarray}
\un Z_{(t) L}(z)
&=&\un z_{(t) 0}
+i\sqrt{2\alpha'} 
e^{- i\gamma_t} 
\sum_{n=0}^{\infty}
\left[
\frac{ \bar {\un \alpha}_{(t) n+1-\epsilon_t} }{\,{n+1-\epsilon_t}} 
z^{-(n+1-\epsilon_t)}
-
\frac{ \un \alpha_{(t) n+\epsilon_t}^\dagger}{\,{n+\epsilon_t}} 
z^{+(n+\epsilon_t)}
\right]
\nonumber\\
\un Z_{(t) R}(\bar z)
&=&
\un z_{(t) 0}
+i\sqrt{2\alpha'} 
e^{+ i\gamma_t} 
%e^{+2 i\gamma_t} 
\sum_{n=0}^{\infty}
\left[ 
\frac{\bar{\un \alpha}_{(t) n+1-\epsilon_t} }{\,{n+1-\epsilon_t}} 
{\bar z}^{-(n+1-\epsilon_t)}
- 
\frac{ \un \alpha_{(t) n+\epsilon_t}^\dagger}{\,{n+\epsilon_t}} 
{\bar z}^{+(n+\epsilon_t)}
\right]
\end{eqnarray}
and
\begin{eqnarray}
\bar{\un Z}_{(t) L}(z)
&=&\un{\bar z}_{(t) 0}
+i\sqrt{2\alpha'} 
e^{+ i\gamma_t} 
\sum_{n=0}^{\infty}
\left[
-\frac{ \bar{\un \alpha}_{(t) n+1-\epsilon_t}^\dagger}{\,{n+1-\epsilon_t}} 
z^{+(n+1-\epsilon_t)}
+
\frac{ \un \alpha_{(t) n+\epsilon_t}}{\,{n+\epsilon_t}} 
z^{-(n+\epsilon_t)}
\right]
\nonumber\\
\bar Z_{(t) R}(\bar z)
&=& \un{\bar z}_{(t) 0}
+i\sqrt{2\alpha'} 
e^{- i\gamma_t} 
%e^{-2i\gamma_t} 
\sum_{n=0}^{\infty}
\left[ 
-\frac{\bar{\un \alpha}_{(t)    n+1-\epsilon_t}^\dagger}{\,{n+1-\epsilon_t}} 
{\bar z}^{+(n+1-\epsilon_t)}
+ 
\frac{\un \alpha_{(t) n+\epsilon_t}}{\,{n+\epsilon_t}} 
{\bar z}^{-(n+\epsilon_t)}
\right]
\end{eqnarray}
The previous fields satisfy the boundary conditions\footnote{
These can also be written as 
$e^{i\gamma_t} \partial \un Z_{(t) L}(x)= 
\frac{1}{\cos \gamma_t} \partial \un Z_{(t)}(x,x) $
when $x>0$
and
$e^{i\gamma_{t-1}} \partial \un Z_{(t) L}(y)= 
\frac{1}{\cos \gamma_{t-1}} \partial \un Z_{(t)}(y,\by) $ when $y<0$.
These expressions are those used to connect the open string operators
when naturally expressed as function of $X(x,\bx)$ to their
expressions as functional of $X_L(x)$.
}
\begin{eqnarray}
e^{i\gamma_t} \partial \un Z_{(t) L} |_x &=& 
e^{-i\gamma_t} \bar\partial \un Z_{(t) R} |_{x} 
~~
x\in \R^+
\\
e^{+i\gamma_{t-1}} \partial \un Z_{(t) L} |_y &=& 
e^{-i\gamma_{t-1}} \bar \partial \un Z_{(t) R} |_y 
~~
y= |y| e^{i\pi}\in \R^-
\end{eqnarray}
where we have defined the phases ($-\frac{\pi}{2}< \gamma_t < \frac{\pi}{2}$)
\begin{eqnarray}
e^{i\gamma_t}= \frac{1 + i B_t }{\sqrt{1+ B_t^2}}
&\rightarrow&
B_t = \tan \gamma_t = 2\pi \alpha'~ q_{(0)}F_{1 2 (0)}
\nonumber\\
e^{i\gamma_{t-1}}= \frac{1 + i B_{t-1} }{\sqrt{1+ B_{t-1}^2}}
&\rightarrow&
B_{t-1} = \tan \gamma_{t-1} = 2\pi \alpha'~ q_{(\pi)} F_{1 2 (\pi)}
\end{eqnarray}
where  $B_{t-1}= 2\pi \alpha'~ q_{(\pi)} F_{1 2 (\pi)}$ and
$B_{t}= 2\pi \alpha'~ q_{(0)}F_{1 2 (0)}$ are the adimensional magnetic
fields which are on the $x<0$ ($\sigma=\pi$) and $x>0$ ($\sigma=0$)
boundaries.
In the field expansion the shift $\epsilon_t$ is given by
%%%%%%%%%%%
\COMMENTOOK{Do bc work for all possible $\gamma_t, \gamma_{t-1}$? Now
  works.}
%%%%%%%%%%%
\begin{equation}
\epsilon_t= 
\left\{
\begin{array}{l c}
\frac{1}{\pi} \left(  \gamma_t -\gamma_{t-1} \right) &  
\gamma_t >\gamma_{t-1}
\\
1+ \frac{1}{\pi} \left(  \gamma_t -\gamma_{t-1} \right) &  
\gamma_t <\gamma_{t-1}
\end{array}
\right.
~~~~
0\le 
\epsilon_t
<1
\end{equation}

The previous operators act on the $\cH_t$ twisted ground state defined by
\begin{equation}
 \bar {\un \alpha}_{(t) n+1-\epsilon_t} |T_t\rangle
=
\un \alpha_{(t) m+\epsilon_t} |T_t\rangle
=
x^2_{(t) 0}  |T_t\rangle
=0
\end{equation}
and have the non vanishing commutation relations
\footnote{Since the annihilator and creator operators have flat indexes
this holds independently of our choice of the taking the metric
diagonal; in particular from
definition of the complex fields we have 
$  (d X^1)^2+ (d X^2)^2= 2 d Z d\bar Z$, i.e. $G_{z \bar z}=1$. }
\begin{eqnarray}
[\un{\bar \alpha}_{(t) n+1-\epsilon_t} , \un{\bar \alpha}_{(t) m+1-\epsilon_t}^\dagger ]
&=&
( n+1-\epsilon_t)
\delta_{n,m} 
~~
n,m \ge 0
\nonumber\\
{}[\un{\alpha}_{(t) n+\epsilon_t} , \un{\alpha}_{(t) m+\epsilon_t}^\dagger ] 
&=&
( n+\epsilon_t)
\delta_{n,m} 
~~
n,m\ge 0
\nonumber\\
{}[\un z_{(t) 0}, \bar{\un z}_{(t) 0}]
&=&
\frac{2 \pi \alpha'}{ \un B_{t} -\un B_{t-1}}
%=
%\frac{2 \pi \alpha'}{ \un{\hat \ff}_{(0)} -\un{\hat \ff}_{(\pi)}}
\end{eqnarray}
Notice that the choice of the definition of the zero modes vacuum is
somewhat arbitrary since they do not change the energy, our choice is
dictated by our gauge choice for the background magnetic field 
$A= B ~x^1 ~d x^2$ which implies the translational invariance 
$X^2 \rightarrow X^2 +\epsilon$
and by the observation that is almost the proper choice in toroidal
compactifications.
The existence of the zero modes imply that the vacuum is degenerate
since all the states
$|T_t, \kappa_t\rangle = e^{i \kappa_t x^1_{(t) 0} }|T_t\rangle$
have exactly the same energy of the vacuum
and therefore there exists a one parameter family of twist fields
(\cite{Pesando:2011yd}) $\sigma_{\epsilon_t, \kappa_t}(x,\bx)$.

Given the previous vacuum definition we have the following twisted Green
functions
\begin{align}
\Gr_{T(t)}^{z z}(z, \bar z; w, \bar w)
& = [Z^{(+)}(z,\bar z), Z^{(-)}(w,\bar w)] |_{an. cont}
=
\frac{\pi \alpha'}{ \un B_{t} -\un B_{t-1}}
\nonumber\\
\Gr_{T(t)}^{\bar z \bar z}(z,\bar z; w, \bar w)
& = [\bar Z^{(+)}(z,\bar z), \bar Z^{(-)}(w,\bar w)] |_{an. cont}
=
-\frac{\pi \alpha'}{ \un B_{t} -\un B_{t-1}}
\nonumber\\
\Gr_{T(t)}^{z \bar z}(z,\bar z; w, \bar w)
& = [Z^{(+)}(z,\bar z), \bar Z^{(-)}(w,\bar w)] |_{an. cont}
\nonumber\\
&=
\frac{\pi \alpha'}{ \un B_{t} -\un B_{t-1}} %~sign(|z|-|w|)
\nonumber\\
&~~
-\frac{\alpha'}{2} 
\left[
\,g_{\epsilon_t}\left( \frac{w}{z} \right)
+\,g_{\epsilon_t}\left( \frac{\bar w}{\bar z} \right)
+ e^{-2i \gamma_t} \,g_{\epsilon_t}\left( \frac{\bar w}{z} \right)
+ e^{ 2i \gamma_t} \,g_{\epsilon_t}\left( \frac{ w}{\bar z} \right)
\right]
\nonumber\\
\Gr_{T(t)}^{\bar z z}(z,\bar z; w, \bar w)
& = [\bar Z^{(+)}(z,\bar z), Z^{(-)}(w,\bar w)]|_{an. cont}
\nonumber\\
&=
-\frac{\pi \alpha'}{ \un B_{t} -\un B_{t-1}} %~sign(|z|-|w|)
\nonumber\\
&~~
-\frac{\alpha'}{2} 
\left[
\,g_{1-\epsilon_t}\left( \frac{w}{z}\right)
+\,g_{1-\epsilon_t}\left( \frac{\bar w}{\bar z}\right)
+ e^{2i \gamma_t} \,g_{1-\epsilon_t}\left( \frac{\bar w}{z}\right)
+ e^{-2i \gamma_t} \,g_{1-\epsilon_t}\left( \frac{ w}{\bar z}\right)
\right]
\end{align}
which can be obtained by analytically continuing their operatorial
expression from $|z| > |w|$ to the whole upper plane in such a way to
preserve the symmetry 
$G^{i j}(z, \bar z; w, \bar w)=G^{i j}(w, \bar w; z, \bar z)$. 
In the previous expressions 
we have defined $\,g_{\nu}(z)%\equiv \,g_{\nu,0}(z)
$ 
as the analytic continuation of 
\begin{equation}
\,g_{\nu,s}(z)= - \sum_{n-\nu>0} \frac{1}{n-\nu} z^{n-\nu}
~~~ 
|z|<1,
~~~
-\pi+2\pi s<\phi=arg(z)\le \pi+ 2 \pi s
.
\end{equation}
in the properly chosen sheet $s$.
Notice that the symmetry of the Green function 
$\Gr^{i j}(z,\bar z; w, \bar w) =\Gr^{j i}(w,\bar w; z, \bar z)$
is not obvious in the zero modes sector, i.e. for the constant terms but 
it holds due to the
$\,g$ transformation property (\cite{Pesando:2011yd})
\begin{eqnarray}
\,g_{\nu,s}(z)
=
\,C_{\nu,s}(\phi)+ \,g_{1-\nu,-s}\left(\frac{1}{z}\right)
,~~~~
\,C_{\nu,s}(\phi)=
\left\{ \begin{array} {cc}
\frac{\pi e^{-i \pi \nu}}{sin \pi\nu} e^{-i 2\pi\nu s}&
2\pi s<\phi<\pi+2\pi s
\\
\frac{\pi e^{+i \pi \nu}}{sin \pi\nu} e^{-i 2\pi\nu s} & 
-\pi+2\pi s<\phi<2\pi s
\end{array}
\right.
\nonumber\\
\label{ghat-an-cont}
\end{eqnarray}
This fact implies that we cannot really completely separate the zero
modes and non zero modes also for the twisted sector as it already
happens for the untwisted one.

For $x>0>y$ and $|y/x|<1$ the previous Green functions become on the
boundary\footnote{
When $|y/x|>1$ we must be more careful since we want to evaluate the
$\,g$ on a cut; for example when $0<x,y$ the expression which is
valid for all ranges is
$
\Gr_{T(t)~bou}^{z \bar z}(x; y)
 =
\frac{\pi \alpha'}{ \un B_{t} -\un B_{t-1}} 
%\nonumber\\
%&~~
-\alpha' ~\cos \gamma_t  
[
e^{-i \gamma_t}
\,g_{\epsilon_t}\left( \frac{y}{x} e^{-i0}\right)
+
e^{i \gamma_t}
\,g_{\epsilon_t}\left( \frac{y}{x} e^{+i0}\right)
].$
In any case we can always use the symmetry property for the Green
functions to reduce the computation 
in the range where we can apply the given expressions.
}
\begin{align}
\label{twisted-bou-green}
\Gr_{T(t)~bou}^{z z}(x; y)
& 
=
\frac{\pi \alpha'}{ \un B_{t} -\un B_{t-1}}
\nonumber\\
\Gr_{T(t)~bou}^{\bar z \bar z}(x; y)
& 
=
-\frac{\pi \alpha'}{ \un B_{t} -\un B_{t-1}}
\nonumber\\
\Gr_{T(t)~bou}^{z \bar z}(x; y)
& =
\frac{\pi \alpha'}{ \un B_{t} -\un B_{t-1}} 
%\nonumber\\
%&~~
-2\alpha' ~\cos \gamma_t  ~\cos \gamma_{t-1} 
~e^{i \gamma_t -i \gamma_{t-1}}
~\,g_{\epsilon_t}\left( \frac{y}{x}\right)
\nonumber\\
\Gr_{T(t)~bou}^{\bar z z}(x; y)
& = 
-\frac{\pi \alpha'}{ \un B_{t} -\un B_{t-1}} 
-2\alpha' ~\cos \gamma_t ~\cos \gamma_{t-1} 
~e^{-i \gamma_t +i \gamma_{t-1}}
~\,g_{1-\epsilon_t}\left( \frac{y}{x}\right)
\end{align}
The other cases can be obtained with the substitution rule 
$ x>0 ~~\cos \gamma_t~e^{i \gamma_t} \leftrightarrow x<0 ~~
\cos \gamma_{t-1}~e^{-i \gamma_{t-1}}$ and the same for $y$ in the
$G^{z \bar z}$ propagator.  For the $G^{\bar z z}$ propagator one
takes the complex conjugate of the previous substitution rule.

\item
In a similar way to any untwisted operator we insert in the amplitude
we associate an auxiliary Hilbert space  $\cH_{a,t_a}$. This Hilbert
space as well as the position where the untwisted vertex is inserted
$x_{a,t_a}$ are better labeled by both a counting label $a=1\dots M$
and a further label $t_a\in\{1,\dots N\}$ 
which specify which is the magnetic field  felt by the untwisted state
(dipole string). 
This could seem irrelevant but it is important in defining the
regularized Green functions (\ref{Green-bou-reg-untwisted}) 
and in computing the non commutative phases.
In the following we will use a lighter notation as
$x_{a,t_a}\rightarrow x_a $ when there is not possibility of
confusion.

On the auxiliary Hilbert space $\cH_{a,t_a}$ act the quantum fields 
\begin{equation}
\un Z_{(a,t_a)}(z,\bar z)
=
\frac{1}{\sqrt{2}}\left( \un X^1_{(a,t_a)} (z,\bar z) 
+i \un X^2_{(a,t_a)} (z,\bar z) \right)
=
\frac{1}{2}( \un Z_{(a,t_a) L}(z) + \un Z_{(a,t_a) R}(\bar z))
\end{equation}
which have expansions
\begin{eqnarray}
\un Z_{(a,t_a) L}
%&=&
%(1- i B_{t_a}) \hat{ \un Z}_{L (0)}+ w_0
%= e^{-i \gamma_{t_a}}  { \un Z}_{L (0)}+ w_0
%\nonumber\\
&=&
e^{-i \gamma_{t_a}} \Bigg(
\un z_{(a,t_a) 0}
-2\alpha'  \un \bp_{(a,t_a)} ~i \ln(z)
+ i \sqrt{2\alpha'} \sum_{n=1}^\infty 
+\frac{\bar { \un \alpha}_{(a,t_a) n} }{\,{n}}  z^{-n}
-\frac{\un \alpha_{(a,t_a) n}^\dagger }{\,{n}}  z^{n}
\Bigg)
% no wilson
%+ w_0
\nonumber\\
%%%%%%%%%%%
\un Z_{(a,t_a) R}
%&=&
%(1+ i B_{t_a}) \hat{ \un Z}_{R (0)} - w_0
%= e^{+i \gamma}  { \un Z}_{R (0)} - w_0
%\nonumber\\
&=&
e^{+i \gamma_{t_a}} \left(
\un z_{(a,t_a) 0}
-2\alpha'  \un \bp_{(a,t_a)} ~i \ln(\bar z)
+i \sqrt{2\alpha'} \sum_{n=1}^\infty
+\frac{\bar { \un \alpha}_{(a,t_a) n} }{\,{n}}  \bar z^{-n}
-\frac{\un \alpha_{(a,t_a) n}^\dagger }{\,{n}}  \bar z^{n}
\right)
%- w_0
\nonumber\\
\end{eqnarray}
%where we have introduced 
%${ \un Z}_{L (0)}= \sqrt{1+\un \ff^2 }\hat{ \un Z}_{L (0)}$
%and for their complex conjugates $\un {\bar
%  Z}=\frac{1}{\sqrt{2}}\left(\un X^1- i \un X^2\right)$  
and
\begin{eqnarray}
\un {\bar Z}_{(a,t_a) L} 
%&=&
%(1+ i \un \ff) \hat{ \bar{ \un Z}}_{L (0)}+ \bar w_0
%= e^{+i \gamma}  { \un {\bar Z}}_{L (0)}+ \bar w_0
%\nonumber\\
&=&
e^{+i \gamma_{t_a}} \left(
\un {\bar z}_{(a,t_a) 0} 
-2\alpha'  \un {\pp}_{(a,t_a)} ~i \ln(z)
+ i \sqrt{2\alpha'} \sum_{n\ne 0} 
-\frac{\bar { \un \alpha}_{(a,t_a) n}^\dagger }{\,{n}}  z^{n}
+\frac{\un \alpha_{(a,t_a) n} }{\,{n}}  z^{-n}
\right)
%+ \bar w_0
\nonumber\\
%%%%%%%%%%%
\un {\bar Z}_{(a,t_a) R}
%&=&
%(1- i \un \ff) \hat{ \un {\bar Z}}_{R (0)} - \bar w_0
%= e^{-i \gamma}  { \un {\bar Z}}_{R (0)} - \bar w_0
%\nonumber\\
&=&
e^{-i \gamma_{t_a}} \left(
\un {\bar z}_{(a,t_a) 0} 
-2\alpha'  \un {\pp}_{(a,t_a)} ~i \ln(\bar z)
+ i \sqrt{2\alpha'} \sum_{n\ne 0} 
-\frac{\bar { \un \alpha}_{(a,t_a) n}^\dagger }{\,{n}}  \bar z^{n}
+\frac{\un \alpha_{(a,t_a) n} }{\,{n}}  \bar z^{-n}
\right)
%- \bar w_0
\nonumber\\
\end{eqnarray}
The previous quantum fields satisfy 
the boundary conditions
\begin{eqnarray*}
e^{+i\gamma_{t_a}} \partial \un Z_{(a,t_a)} |_x &=& 
e^{-i\gamma_{t_a}} \bar\partial \un Z_{(a,t_a)} |_{x} 
~~
x\in \R^+
\\
e^{+i\gamma_{t_a}} \partial \un Z_{(a,t_a)} |_y &=& 
e^{-i\gamma_{t_a}} \bar\partial \un Z_{(a,t_a)} |_y 
~~
y= |y| e^{i\pi}\in \R^-
\end{eqnarray*}
where we have defined the angle $\gamma_{t_a}$, in a similar way for the twisted
scalar (dicharged string), as
\begin{equation}
e^{i \gamma_{t_a}}= \frac{1+ i B_{t_a}}{\sqrt{1+ B_{t_a}}}
\Rightarrow
B_{t_a} = \tan \gamma_{t_a}
,~~~~
-\frac{\pi}{2}<\gamma_{t_a}<\frac{\pi}{2}
\end{equation}

%%%%%%%%%%%%%%%%%%%%%%% COMMENTO
\COMMENTO{
In the following we need the background matrices in flat real coordinates
which we define as
%%%%%%%%%%
\COMMENTOO{
Better in flat complex coords.
}
%%%%%%%%%%%
\begin{align}
\un \cE_{(t_a)}
=
\left(\begin{array}{c c}
1 & B_{t_a} \\ 
-B_{t_a} & 1 
\end{array}\right),
\end{align}
from which we define the open string background
\begin{align}
\un \cG^{-1}_{(t_a)}
&=
\un \cG^{-1}(B_{t_a})
=
\frac{1}{1+ B_{t_a}^2}
\left(\begin{array}{c c}
1 & 0 \\ 
0 & 1 
\end{array}\right)
=\cos^2 \gamma_{t_a} ~\uno_2
,
\nonumber\\
\un \Theta_{(t_a)} 
&= 
\frac{1}{1+ B_{t_a}^2}
\left(\begin{array}{c c}
0 & B_{t_a} \\ 
-B_{t_a} & 0 
\end{array}\right)
=\cos \gamma_{t_a} \sin \gamma_{t_a} ~\epsilon_2
\end{align}
}
%%%%%%%%%%%%%%%%%%%%%%% COMMENTO

The creation and destruction operators act on the dipole ground state
defined by
\begin{equation}
\bar \alpha_{(a,t_a) n} |0_{(a,t_a)}\rangle
=
 \alpha_{(a,t_a) n} |0_{(a,t_a)}\rangle
=
\bp_{(a,t_a)} |0_{(a,t_a)}\rangle
=
\pp_{(a,t_a) } |0_{(a,t_a)}\rangle
=
0
\end{equation}
and have non trivial commutation relations
\begin{eqnarray}
\label{comm-rel-dicharged}
[\un z_{(a,t_a) 0}, \bar{\un z}_{(a,t_a) 0} ]&=& 2\pi \alpha' B_{t_a}
\nonumber\\
{}[\un z_{(a,t_a) 0}, \un{\pp}_{(a,t_a)} ]&=& i
\nonumber\\
{}[\un \alpha_{(a,t_a) n} ,{\un \alpha}^\dagger_{(a,t_a) m}] &=&  n \delta_{m,n}
\nonumber\\
{}[\un \bar \alpha_{(a,t_a) n} ,\bar {\un \alpha}^\dagger_{(a,t_a) m}] &=& n \delta_{m,n}
\end{eqnarray}

The normal ordering is the usual one but it worth noticing that in the
zero modes sector is defined as
%$
%:e^{i k_i X^i_{(a) zm}(x, \bx)}: = e^{i k_i x^i_{(a)}} e^{2\alpha' k_i p^i_{(a)} ~\ln(|x|)}
%$
%or in complex notation
\begin{equation}
:e^{i (\bar k Z_{(a) zm}+ k \bar Z_{(a) zm})(x, \bx)}: = 
\left\{
\begin{array}{c r}
e^{i \cos\gamma_t (\bar k z_{(a) 0}+ k \bar z_{(a)0})} 
~e^{2\alpha' \ln(|x|) \cos\gamma_t (\bar k \bp_{(a) 0}+ k \pp_{(a)0})}
& x>0
\\
e^{i \cos\gamma_t (\bar k \hat z_{(a) 0}+ k \hat {\bar z}_{(a)0})} 
~e^{2\alpha' \ln(|x|) \cos\gamma_t (\bar k \bp_{(a) 0}+ k \pp_{(a)0})}
&x <0
\end{array}
\right.
.
\end{equation}
with $ \hat z_{(a) 0}=   z_{(a) 0} - i 2\pi \alpha' \tan\gamma_t~\bp $
which have the property that their commutation relations are the
opposite of the $ z_{(a) 0}$ ones. 
Finally the untwisted Green functions in a magnetic background
$B_{t_a}$ are given by ($0<  arg(z-\bar w)<\pi$) 
\begin{align}
\Gr_{U (t_a)}^{z z}(z,\bar z, w, \bar w)
&=
\Gr_{U (t_a)}^{\bar z \bar z}(z,\bar z, w, \bar w)
=0
\nonumber\\
\Gr_{U (t_a)}^{z \bar z}(z,\bar z, w, \bar w)
& = [Z^{(+)}(z,\bar z), \bar Z^{(-)}(w,\bar w)] |_{an.cont}
\nonumber\\
&=
+
\oh \pi \alpha'\sin(2\gamma_{t_a})
-\alpha' 
\left[
\ln |z-w| + \cos(2\gamma_{t_a})  \ln|z- \bar w|
+\sin(2\gamma_{t_a}) arg(z-\bar w)
\right]
\nonumber\\
\Gr_{U (t_a)}^{\bar z z}(z,\bar z, w, \bar w)
& = [\bar Z^{(+)}(z,\bar z), Z^{(-)}(w,\bar w)] |_{an.cont}
\nonumber\\
&=
-
\oh \pi \alpha'\sin(2\gamma_{t_a})
-\alpha' 
\left[
\ln |z-w| + \cos(2\gamma_{t_a})  \ln|z- \bar w|
-\sin(2\gamma_{t_a}) arg(z-\bar w)
\right]
\end{align}
The constant terms can be obtained by rewriting  
\begin{equation}
 z_{(a) 0}=   z_{(a) 0 0} +i  \pi \alpha' \tan\gamma_t~\bp 
\end{equation}
so that $[ z_{(a) 0 0},  \bar z_{(a) 0 0}]=0$ an considering the
additional term proportional to $\bp$ 
coming from this rewriting as belonging to 
$ Z^{(+)}(z,\bar z)$.
Notice however once again that the constant terms are needed to ensure the
symmetry $\Gr^{i j}(z,\bar z; w, \bar w) =\Gr^{j i}(w,\bar w; z, \bar z)$.

The previous Green functions become on the boundary $z=x, w=y\in\R$ 
\footnote{
When using these Green functions in eq. (\ref{PI_untwisted_1}) in
absence of twist fields we recover
the results from the operatorial formalism.
}
\begin{align}
\label{dipole-Gr-bou}
\Gr_{U (t_a),~bou}^{z \bar z}(x; y )
&=
\oh \pi \alpha'\sin(2\gamma_{t_a})
-2\alpha' 
\left[
 \cos^2(\gamma_{t_a})  \ln|x- y|
+\oh \sin(2\gamma_{t_a}) arg(x-\bar y)
\right]
\nonumber\\
\Gr_{U (t_a),~bou}^{\bar z z}(x; y)
&=
-
\oh \pi \alpha'\sin(2\gamma_{t_a})
-2\alpha' 
\left[
\cos^2(\gamma_{t_a})  \ln|x- \bar y|
-\oh \sin(2\gamma_{t_a}) arg(x-\bar y)
\right]
\end{align}
From these expressions we can read the open string metric
$\cG^{z \bar z}_{(t)}= \cos^2(\gamma_t)$ and the non commutativity parameter 
$\Theta^{z \bar z}_{(t)}= \oh \sin(2\gamma_t)$, we can also read the
$\R^2$ vielbein
$\cV^{\underline {z} \bar z}_{(t)}=\cV^{\underline {\bar z} z}_{(t)}
= \frac{1}{ \cos(\gamma_t)}$ where $\underline {z},\underline {\bar
  z}$ are the flat indexes which are also implicit in the creation and
destruction operators.

\item
We define 
the boundary Green function regularized by the untwisted Green function
for a background $B_{t_a}$ as 
%%%%%%%%%%
\COMMENTOO{ This is in the trivial conformal gauge.}
%%%%%%%%%%
\begin{align}
\label{Green-bou-reg-untwisted}
\Gr_{bou,~reg~ U (t_a) }^{i j}(x; y; \{x_v\})
=
&
%\oh\left(
\Gr^{i j}_{bou}(x; y; \{x_v\})
-
\Gr_{U (t_a),~bou}^{i j}(x; y)
%\right)
%\nonumber\\
%&
%+
%\oh\left(
%\Gr^{i j}_{bou}(y; x; \{x_v\})
%-
%\Gr_{U (t_a),~bou}^{i j}(y; x)
%\right)
~~~~
x,y\in\R
\end{align}
where
$\Gr_{U (t_a),~bou}^{i j}(x,y)$ 
are defined in eq.s (\ref{dipole-Gr-bou}). The choice of the
background $B_t$  in the regularization  would seem arbitrary but it
is not since these regularized Green functions (and their derivatives) 
enter only where an untwisted dipole state is emitted and this is 
on a well defined interval of the boundary.
%In the main expression we will also use the analytical continuation
%from $x, y\in\R$ to $z, w\in\C$. 

We also define
the analogous twisted boundary Green function regularized by the twisted Green
function at the twist insertion point
$t$  as\footnote{The symmetrization is because we a symmetric function
in $x\leftrightarrow y$, i.e. independent on the way we take the limit
$x>y$ or $y>x$.}
 %($x>y$, i.e $x=x_a^+$ and $y=x_a$ for the first expression)
\begin{align}
\label{Green-bou-reg-T}
\Gr^{i j}_{bou,~reg~T(t)}(x, y; \{x_v\})
&=
\Gr^{i j}_{bou}(x, y; \{x_v\})
-
\Gr^{i j}_{T(t)~bou}(x, y; \{x_0=x_t, x_\infty=\infty\})
\end{align}
where $\Gr^{i j}_{t~bou}$  are given in eq.s (\ref{twisted-bou-green})
with the substitution $ \frac{y}{x} \rightarrow \frac{y-x_t}{x-x_t}$.
\end{itemize}

Given the previous building blocks the main formula is given by
($x_t\ne x_a~~~\forall t,a$)
%%%%%%%%%%%
\COMMENTOO{
1) Better written using dipole aux fields or normal aux
  fields?
The dipole bck changes along the boundary hence dipole aux fields must
be used.\\
2) Is it better using left moving fields so T duality is easier?
The operator algebra depends however on the frame in the $z_0$ part only.
Moreover with $L$ fields the analytical continuation is obvious.
\\
3) $\hat Z_L$ is $Z_L$ without the phase and similarly all the others.
\\
} 
%%%%%%%%%%%
\begin{align*}
&
\langle V_{N+M}(\{x_t\}_{t=1,\dots N}; \{x_{a,t_a}\}_{a=1,\dots M})|
=
C(x_1,\dots x_N)
\nonumber\\
&
\prod_{a=1}^M \langle 0_{(a) a}, z_{(a) 0 0}=\bar z_{(a) 0 0}=0|
\prod_{t=1}^N \langle T_{(t)}, x^1_{(t)}=0|
\delta( i \sum_a (\alpha_{(a)  0} - \bar \alpha_{(a)  0} )
+
i \sum_t ( z_{(t)  0} - \bar z_{(t)  0} )
)
\nonumber\\
&
\prod_a
\exp\Big\{
-\frac{1}{4\alpha'}
\alpha_{(a) 0}^2 
~\cV_{(t_a)~\underline{ z} \bar z}^2
~\Gr^{\bar z \bar z}_{bou,~reg~U(t_a)}(x; y; \{x_v\})
-
\frac{1}{4\alpha'}
\bar \alpha_{(a) 0}^2 
~\cV_{(t_a)~\underline{\bar z} z}^2
~\Gr^{z z}_{bou,~reg~U(t_a)}(x; y; \{x_v\})
\nonumber\\
&
%%%%%%%
\phantom{\prod_a\exp\Big\{}
%%%%%%%
-\frac{1}{2\alpha'}
 \sum_{n,m=0}^\infty \alpha_{(a) n} ~\bar \alpha_{(a) m}
~\cV_{(t_a)~\underline{\bar z} z}\cV_{(t_a)~\underline{ z} \bar z }
~\frac{ \partial^n_x }{ n! } 
~\frac{ \partial^m_y }{ m!}
~\Gr^{z \bar z}_{bou,~reg~U(t_a)}(x; y; \{x_v\})
\Big\}
\Big |_{x=y=x_a}
\nonumber\\
%%%%%%%%%%
&
\prod_{t }
\exp\Big\{
%%%%%%%% PIU CHIARO CON x^2
%\oh d_{(t) 0}^2   ~\Gr^{\bar z \bar z}_{bou,~reg~T(t)}(x; y; \{x_v\})
%+
%\oh \bar d_{(t) 0}^2   ~\Gr^{z z}_{bou,~reg~T(t)}(x; y; \{x_v\})
%\nonumber\\
%&\hspace{3em}
%+
% d_{(t) 0} ~\bar  d_{(t) 0}  ~\Gr^{z \bar z}_{bou,~reg~T(t)}(x; y; \{x_v\})
\oh \left( \frac{\tan \gamma_t - \tan \gamma_{t-1}}{2\pi \alpha'} 
{ x_{(t) 0~i=2} }{ } \right)^2 ~\Gr^{2 2}_{bou,~reg~T(t)}(x; y; \{x_v\})
\nonumber\\
&\hspace{3em}
-\frac{1}{2\alpha'}
 \sum_{n,m=1}^\infty 
\frac{ \bar \alpha_{(t) n}}{n-\epsilon_t} 
\frac{ \alpha_{(t) m}}{m-1+\epsilon_t} 
~\cV_{(t)~\underline{z} \bar z}\cV_{(u)~\underline{\bar  z} z }
\nonumber\\
&\hspace{7em}
~\frac{ \partial^{n-1}_x }{ (n-1)!}
~\frac{ \partial^{m-1}_y }{ (m-1)!}
\Big[(x-x_t)^{1-\epsilon_t} (y-x_t)^{\epsilon_t}
~\partial_x \partial_y
\Gr^{z \bar z}_{bou,~reg~T(t)}(x, y; \{x_v\})
\Big]
\Big\}
\Big|_{x=y= x_t}
%\nonumber\\
\end{align*}
%%%%%%%%%%%%%
\begin{align*}
&
\prod_{a<b}
\exp\Big\{
-
\frac{1}{2\alpha'}
\alpha_{(a) 0 } ~\alpha_{(b) 0}
~\cV_{(t_a)~\underline{\bar z} z}\cV_{(t_b)~\underline{ \bar z} z }
~\Gr^{z z}_{bou}(x ; y; \{x_v\})
-
\frac{1}{2\alpha'}
\bar \alpha_{(a) 0 } ~\bar \alpha_{(b) 0}
~\cV_{(t_a)~\underline{z} \bar z}\cV_{(t_b)~\underline{z} \bar z }
~\Gr^{\bar z \bar z}_{bou}(x ; y; \{x_v\})
\nonumber\\
&
\phantom{ \prod_{ a < b} \exp\Big\{ }
-\frac{1}{2\alpha'}
\sum_{n,m=0}^\infty 
\alpha_{(a) n } ~\bar \alpha_{(b) m}
~\cV_{(t_a)~\underline{\bar z} z}\cV_{(t_b)~\underline{ z} \bar z }
~\frac{ \partial^n_x }{ n! } 
~\frac{ \partial^m_y }{ m!}
~\Gr^{z \bar z}_{bou}(x ; y; \{x_v\})
\Big\}
\nonumber\\
&
%%%%%%%
\phantom{\prod_{a<b} \exp\Big\{ }
%%%%%%%
-\frac{1}{2\alpha'}
\sum_{n,m=0}^\infty \bar \alpha_{(a) n } ~\alpha_{(b) m}
~\cV_{(t_b)~\underline{\bar z} z}\cV_{(t_a)~\underline{ z} \bar z }
~\frac{ \partial^n_x }{ n! } 
~\frac{ \partial^m_y }{ m!}
~\Gr^{\bar z z}_{bou}(x ; y; \{x_v\})
\Big\}
\Big|_{x=x_a, y=x_b}
%\nonumber\\
%\label{Final-Reggeon-N+M}
\end{align*}
%%%%%%%%%%
\begin{align*}
%%%%%%%%%%%
%\phantom{}
&
\prod_{t < u}
\exp\Big\{
%d_{(t) 0} ~d_{(u) 0}   ~\Gr^{\bar z \bar z}_{bou}(x_t, x_u; \{x_v\})
%+
%\bar d_{(t) 0}  ~\bar d_{(u) 0} ~\Gr^{z z}_{bou}(x_t, x_u; \{x_v\})
%\nonumber\\
%&\hspace{3em}
%+
%d_{(t) 0} ~\bar  d_{(u) 0}  ~\Gr^{\bar z z}_{bou}(x_t, x_u; \{x_v\})
%+
%\bar d_{(t) 0} ~d_{(u) 0}  ~\Gr^{z \bar z}_{bou}(x_t, x_u; \{x_v\})
%
\left( \frac{\tan \gamma_t - \tan \gamma_{t-1}}{2\pi \alpha'} 
{ x_{(t) 0~i=2} }{ } \right) 
\left( \frac{\tan \gamma_u - \tan \gamma_{u-1}}{2\pi \alpha'} 
{ x_{(u) 0~i=2} }{ } \right) 
~\Gr^{2 2}_{bou}(x; y; \{x_v\})
\nonumber\\
&\hspace{3em}
-\frac{1}{2\alpha'}
 \sum_{n,m=1}^\infty 
\frac{ \bar \alpha_{(t) n}}{n-\epsilon_t} 
\frac{ \alpha_{(u) m}}{m-1+\epsilon_u} 
~\cV_{(t)~\underline{z} \bar z}\cV_{(u)~\underline{\bar  z} z }
\nonumber\\
&\hspace{7em}
~\frac{ \partial^{n-1}_x }{ (n-1)!}
~\frac{ \partial^{m-1}_y }{ (m-1)!}
\left[(x-x_t)^{\epsilon_t} (y-x_u)^{1-\epsilon_u}
\partial_x \partial_y \Gr^{\bar z z}_{bou}(x, y; \{x_v\})
\right]
%\Big\}
\nonumber\\
&\hspace{3em}
-\frac{1}{2\alpha'}
 \sum_{n,m=1}^\infty 
\frac{  \alpha_{(t) n}}{n-1+\epsilon_t} 
\frac{ \bar \alpha_{(u) m}}{m-\epsilon_u} 
~\cV_{(t)~\underline{\bar z}  z}\cV_{(u)~\underline{ z} \bar z }
\nonumber\\
&\hspace{7em}
~\frac{ \partial^{n-1}_x }{ (n-1)!}
~\frac{ \partial^{m-1}_y }{ (m-1)!}
\left[(x-x_t)^{1-\epsilon_t} (y-x_u)^{\epsilon_u}
\partial_x \partial_y \Gr^{z \bar z }_{bou}(x, y; \{x_v\})
\right]
\Big\}
\Big|_{x=x_t,y=x_v}
%\nonumber\\
\end{align*}
%%%%%%%%%%%%
\begin{align}
&
\prod_{t, a}
\exp\Big\{
%%%%%%%%%% GIA NEL TERMINE DOPO
%d_{(t) 0} ~c_{(a) 0}  ~\Gr^{\bar z \bar z}_{bou}(x, y; \{x_v\})
%+
%\bar d_{(t) 0} ~\bar c_{(a) 0}  ~\Gr^{z z}_{bou}(x, y; \{x_v\})
%\nonumber\\
%&
%\hspace{3em}
%+
%%%%%%%%%%%% RISCRIVO CON x^2
%\sum_{m=0}^\infty
%d_{(t) 0} ~\bar c_{(a) m}  ~\partial^m_y \Gr^{\bar z z}_{bou}(x, y; \{x_v\})
%+
%\bar d_{(t) 0} ~c_{(a) m}  ~\partial^m_y \Gr^{z \bar z}_{bou}(x, y; \{x_v\})
-
\frac{1}{2\alpha'}
\left( \frac{\tan \gamma_t - \tan \gamma_{t-1}}{\pi } 
\frac{ x_{(t) 0~i=2} }{ \sqrt{2\alpha'}} \right) 
\sum_{m=0}^\infty
~\alpha_{(a) m}
\cV_{(t_a)~\underline{\bar  z} z }
~\frac{ \partial^m_y }{ m!}
\Gr^{2 z}_{bou}(x, y; \{x_v\})
\nonumber\\
&
-
\frac{1}{2\alpha'}
\left( \frac{\tan \gamma_t - \tan \gamma_{t-1}}{\pi } 
\frac{ x_{(t) 0~i=2} }{ \sqrt{2\alpha'}} \right) 
\sum_{m=0}^\infty
~\bar \alpha_{(a) m}
\cV_{(t_a)~\underline{ z} \bar z }
~\frac{ \partial^m_y }{ m!}
\Gr^{2 \bar z}_{bou}(x, y; \{x_v\})
\nonumber\\
&
\hspace{3em}
-\frac{1}{2\alpha'}
 \sum_{n=1,m=0}^\infty 
\frac{ \bar \alpha_{(t) n}}{n-\epsilon_t} 
~\alpha_{(a) m}
~\cV_{(t)~\underline{z} \bar z}\cV_{(t_a)~\underline{\bar  z} z }
~\frac{ \partial^{n-1}_x }{ (n-1)!}
~\frac{ \partial^m_y }{ m!}
\left[(x-x_t)^{\epsilon_t} 
\partial_x \Gr^{\bar z z}_{bou}(x, y; \{x_v\})
\right]
\nonumber\\
&
\hspace{3em}
-\frac{1}{2\alpha'}
 \sum_{n=1,m=0}^\infty 
\frac{ \alpha_{(t) n}}{n-1+\epsilon_t} 
~\bar \alpha_{(a) m}
~\cV_{(t)~\underline{\bar z}  z}\cV_{(t_a)~\underline{ z} \bar z }
~\frac{ \partial^{n-1}_x }{ (n-1)!}
~\frac{ \partial^m_y }{ m!}
\left[(x-x_t)^{1-\epsilon_t} 
\partial_x \Gr^{\bar z z}_{bou}(x, y; \{x_v\})
\right]
\Big\}
\Big|_{x=x_t, y=x_a}
\label{Final-Reggeon-N+M}
\end{align}
where the operator indexes are raised an lowered using the flat metric
while Green function indexes are raised an lowered using $\R^2$ metric.
The previous expression can also be written without using the
auxiliary operators as a more conventional generating function.
In order to do so we introduce the auxiliary parameters $d_{(t) n}$,
$\bar d_{(t) n}$ and $ c_{(a) n}$ and $\bar c_{(a) n}$ which roughly
correspond to 
$\alpha_{(t) n+1-\epsilon}$, $\bar \alpha_{(t) n-\epsilon}$ and
$\alpha_{(a) n}$, $\bar \alpha_{(a) n}$
(see eq.s (\ref{d-realization}) and (\ref{c-realization}) 
for a precise mapping)
of the previous expression.
Then we can write the generating function as
\COMMENTOO{ 1) Check that $\oh$ becomes $1$ when summing 2) Are there
  the diagonal pieces with $[\Gr]_{reg}$?}
\begin{align*}
&
\cV_{N+M}(\{d_{(t)}\}_{t=1,\dots N}; \{c_{(a)}\}_{a=1,\dots M}
;\{x_t\}_{t=1,\dots N}; \{x_{a,t_a}\}_{a=1,\dots M})
=
\nonumber\\
=&
\delta\left( Re\left(\sum_t d_{(t) 0} +\sum c_{(a) 0})\right)\right)
~C(x_1,\dots x_N)
\nonumber\\
%%%%%%%%%
&
\prod_{a }
\exp\Big\{
\oh
c_{(a) 0}^2 ~\Gr^{\bar z \bar z}_{bou,~reg~U(t_a)}(x; y; \{x_v\})
+
\oh
\bar c_{(a) 0}^2 ~\Gr^{z z}_{bou,~reg~U(t_a)}(x; y; \{x_v\})
\nonumber\\
&
\phantom{  \prod_a \exp\Big\{ }
 \sum_{n,m=0}^\infty c_{(a) n} ~\bar c_{(a) m}
~\partial^n_x
\partial^m_y
\Gr^{z \bar z}_{bou,~reg~U (t_a)}(x; y; \{x_v\})
\Big\}
\Big|_{x=y=x_a} 
\nonumber\\
%%%%%%%%%%
&
\prod_{t }
\exp\Big\{
\oh d_{(t) 0}^2   ~\Gr^{\bar z \bar z}_{bou,~reg~T(t)}(x; y; \{x_v\})
+
\oh \bar d_{(t) 0}^2   ~\Gr^{z z}_{bou,~reg~T(t)}(x; y; \{x_v\})
\nonumber\\
&\hspace{3em}
+
 d_{(t) 0} ~\bar  d_{(t) 0}  ~\Gr^{z \bar z}_{bou,~reg~T(t)}(x; y; \{x_v\})
\nonumber\\
&\hspace{3em}
+ \sum_{n,m=1}^\infty \bar d_{(t) n} ~ d_{(t) m}
~\partial^{n-1}_x
\partial^{m-1}_y
\Big[(x-x_t)^{1-\epsilon_t} (y-x_t)^{\epsilon_t}
~\partial_x \partial_y
\Gr^{z \bar z}_{bou,~reg~T(t)}(x, y; \{x_v\})
\Big]
\Big\}
\Big|_{x=y= x_t}
%\nonumber\\
\end{align*}
\begin{align*}
&
\prod_{ a < b}
\exp\Big\{
\bar c_{(a) n} ~ \bar c_{(b) m}
~\Gr^{z z}_{bou}(x ; y; \{x_v\})
+
 c_{(a) n} ~ c_{(b) m}
~\Gr^{\bar z \bar z}_{bou}(x ; y; \{x_v\})
%\Big\}
\nonumber\\
&
\phantom{ \prod_{ a < b} \exp\Big\{ }
+ \sum_{n,m=0}^\infty \bar c_{(a) n} ~ c_{(b) m}
%\int_{\partial H} d x \int_{\partial H} d y
~\partial^n_x%|_{x=x_a} %\delta_{reg}(x-x_a)
~\partial^m_y%|_{y=x_b} %\delta_{reg}(y-x_a)
~\Gr^{z \bar z}_{bou}(x ; y; \{x_v\})
\Big\}
\nonumber\\
&
\phantom{ \prod_{ a < b} \exp\Big\{ }
+ \sum_{n,m=0}^\infty c_{(a) n} ~ \bar c_{(b) m}
%\int_{\partial H} d x \int_{\partial H} d y
~\partial^n_x%|_{x=x_a} %\delta_{reg}(x-x_a)
~\partial^m_y%|_{y=x_b} %\delta_{reg}(y-x_a)
~\Gr^{\bar z z}_{bou}(x ; y; \{x_v\})
\Big\}
\Big|_{x=x_a,y=x_b}
\end{align*}
\begin{align*}
%%%%%%%%%%%
%\phantom{}
&
\prod_{t < u}
\exp\Big\{
d_{(t) 0} ~d_{(u) 0}   ~\Gr^{\bar z \bar z}_{bou}(x_t, x_u; \{x_v\})
+
\bar d_{(t) 0}  ~\bar d_{(u) 0} ~\Gr^{z z}_{bou}(x_t, x_u; \{x_v\})
\nonumber\\
&\hspace{3em}
+
d_{(t) 0} ~\bar  d_{(u) 0}  ~\Gr^{\bar z z}_{bou}(x_t, x_u; \{x_v\})
+
\bar d_{(t) 0} ~d_{(u) 0}  ~\Gr^{z \bar z}_{bou}(x_t, x_u; \{x_v\})
\nonumber\\
&\hspace{3em}
+
 \sum_{n,m=1}^\infty d_{(t) n} ~\bar d_{(u) m}
~\partial^{n-1}_x
\partial^{m-1}_y
\left[(x-x_t)^{\epsilon_t} (y-x_u)^{1-\epsilon_u}
\partial_x \partial_y \Gr^{\bar z z}_{bou}(x, y; \{x_v\})
\right]
%\Big\}
\nonumber\\
&
\hspace{3em}
+
 \sum_{n,m=1}^\infty \bar d_{(t) n} ~d_{(u) m}
~\partial^{n-1}_x
\partial^{m-1}_y
~\left[(x-x_t)^{1-\epsilon_t} (y-x_u)^{\epsilon_u}
\partial_x \partial_y \Gr^{z \bar z }_{bou}(x, y; \{x_v\})
\right]
\Big\}
\Big|_{x=x_t,y=x_v}
%\nonumber\\
\end{align*}
\begin{align}
&
\prod_{t, a}
\exp\Big\{
%%%%%%%%%%% GIA COMPRESO NEL TERMINE DOPO
%d_{(t) 0} ~c_{(a) 0}  ~\Gr^{\bar z \bar z}_{bou}(x, y; \{x_v\})
%+
%\bar d_{(t) 0} ~\bar c_{(a) 0}  ~\Gr^{z z}_{bou}(x, y; \{x_v\})
%\nonumber\\
%&
%\hspace{3em}
%+
\sum_{m=0}^\infty
d_{(t) 0} ~\bar c_{(a) m}  ~\partial^m_y \Gr^{\bar z z}_{bou}(x, y; \{x_v\})
+
\bar d_{(t) 0} ~c_{(a) m}  ~\partial^m_y \Gr^{z \bar z}_{bou}(x, y; \{x_v\})
\nonumber\\
&
\hspace{3em}
+
 \sum_{n=1,m=0}^\infty 
d_{(t) n} ~\bar c_{(a) m}
~\partial^{n-1}_x
\partial^m_y
\left[(x-x_t)^{\epsilon_t} 
\partial_x \Gr^{\bar z z}_{bou}(x, y; \{x_v\})
\right]
\nonumber\\
&
\hspace{3em}
+
 \sum_{n=1,m=0}^\infty \bar d_{(t) n} ~c_{(a) m}
~\partial^{n-1}_x
\partial^m_y
\left[(x-x_t)^{1-\epsilon_t} 
\partial_x \Gr^{\bar z z}_{bou}(x, y; \{x_v\})
\right]
\Big\}
\Big|_{x=x_t, y=x_a}
\label{Final-Reggeon-N+M-using-cd}
\end{align}

Notice that all the previous expressions are meaningful because of the
behavior of the Green functions
\COMMENTOOK{
ZERO MODES! They enter in the Green functions as it can be verified
from $N=2$ and arbitrary $M$ using the generalized SDS vertex.
}
\begin{align}
\label{Green-asymptotic}
\Gr^{z z}_{bou}(x; y; \{x_v\}) 
&= const
\nonumber\\
\Gr^{\bar z \bar z}_{bou}(x; y; \{x_v\}) &
= const
\nonumber\\
\Gr^{z \bar z }_{bou}(x; y; \{x_v\}) 
=
\Gr^{\bar z  z}_{bou}(y; x; \{x_v\}) 
&{\sim}_{x\rightarrow x_t} 
const+ (x-x_t)^{\epsilon_t}
[ g^{z \bar z }_0(y; \{x_v\}) +O(x-x_t)]
\nonumber\\
\Gr^{z \bar z}_{bou}(x; y; \{x_v\})
=
\Gr^{\bar z z}_{bou}(y; x; \{x_v\})  
&\sim_{y\rightarrow x_u} 
const+ (y-x_u)^{1-\epsilon_u}
[ g^{z \bar z }_0(x; \{x_v\}) +O(y-x_u)]
\nonumber\\
\Gr^{\bar z z}_{bou}(x; y; \{x_v\})
=
\Gr^{z \bar z}_{bou}(y; x; \{x_v\})  
&\sim_{x\rightarrow y; x,y\in(x_t,x_{t+1})} 
const -2 \alpha' \cos^2 \gamma_t log|x-y|
+O(x-y)
\end{align}
where $g^{z \bar z }_0$ and $ g^{z \bar z }_0$ are some functions of
the given variables and
the last line is strictly speaking true when $x\rightarrow y $ but not
at the same
time when $x\rightarrow x_t $ and $y\rightarrow x_t$.
\COMMENTO{
\footnote{
The apparent different result obtained by first taking $y\rightarrow
x_t$ and then $x\rightarrow x_t $ is due to the fact if we consider
the asymptotic behavior of the Green function with respect to both $x$
and $y$ we get a leading order
$(y-x_t)\left(\frac{x-x_t}{y-x_t}\right)^{\epsilon_t}$
in which the term between parenthesis is of order $1$  and  therefore we
cannot limit ourselves to the leading order 
but have to sum all the other terms in the expansion.
}.
}
It is anyhow true that $(x-x_t)^{1-\epsilon_t} (y-x_t)^{\epsilon_t}
~\partial_x \partial_y\Gr^{z \bar z}_{bou,~reg~T(t)}(x, y; \{x_v\})$
is well defined for $x=y=x_t$ as discussed in the appendix \ref{Green-x-y-xt}.

The rest of the paper is organized in the following way: in the next
section we make some examples of the use of the previous formulae and
we clarify the operator to state mapping we use in the twisted sector.
In section \ref{sect:untwisted} we derive the previous formulae for
the case with non excited twisted matter and finally in section
\ref{sect:twisted} we consider excited twisted matter.

\section{Examples}
We want now apply the main formulae stated in the previous section to
some examples while elucidating the nature of excited twisted states.

We start from the simplest example and then move to some more complex
ones while in
appendix \ref{app:N=2-check} we check the $N=2$ not excited states
and $M$ tachyons amplitude against the result found in (\cite{Pesando:2011yd}).

\subsection{Example 1: $N$ not excited twisted states}
From the operator to auxiliary state map
\begin{equation}
\sigma_{\epsilon_t, \kappa_t}(x_t,\bar x_t) \leftrightarrow 
|T_t,\kappa_t\rangle
=\lim_{x\rightarrow 0} \sigma_{\epsilon_t,\kappa_t}(x, \bar x)  |0_{SL}\rangle
\end{equation}
we deduce that
\begin{align}
&
\langle \sigma_{\epsilon_1, \kappa_1}(x_1, \bar x_1)\dots
\sigma_{\epsilon_N, \kappa_N}(x_N, \bar x_N)\rangle
\nonumber\\
&
=
\langle V_{N+0}(\{x_t\}_{t=1,\dots N}|
\prod_t |T_t,\kappa_t\rangle
\nonumber\\
&
=
 \delta(\sum_t \kappa_t)
e^{-\oh \sum_{t} \kappa_t^2  ~G^{2 2}_{bou,~reg~T(t)}(x_t; x_t)}
e^{-\oh \sum_{u,t} \kappa_t \kappa_u ~G^{2 2}_{bou}(x_t; x_u)}
~
C(x_1,\dots x_N)
\end{align}
where the phases proportional  to $\kappa$ probably vanish as they do
in the $N=2$ case but this must be checked with an explicit
computation of the Green function which can, in principle, be
extracted from (\cite{Abel:2003yx}) after T-dualizing.
The same computation can be performed using the more conventional
generating function as
\begin{align}
&
\langle \sigma_{\epsilon_1, \kappa_1}(x_1, \bar x_1)\dots
\sigma_{\epsilon_N, \kappa_N}(x_N, \bar x_N)\rangle
=
\cV_{N+M} \prod_t 
e^{i \kappa_t  
\frac{ \stackrel{\leftarrow}{\partial} }{\partial d_{(t) 0}^2 }
}
\Big|_{c=0; d=0}
.
\end{align}

\subsection{Example 2: $N$ not excited twisted  and $2$ untwisted states}
Similarly to what done in the previous example from the maps
\begin{align}
&
\sigma_{\epsilon_t}(x_t,\bar x_t) 
\leftrightarrow 
|T_t\rangle
=\lim_{x\rightarrow 0} \sigma_{\epsilon_t}(x, \bar x) |0_{SL}\rangle
\nonumber\\
&
\partial X^z(y_a,\bar y_a)
\leftrightarrow 
-i \sqrt{2\alpha'} \cos\gamma_{t_a}~\alpha^\dagger_{(a) 1}| 0_{(a)}\rangle
=\lim_{y\rightarrow 0^+} \partial Z_{(a)}(y, \bar y)  |0_{SL}\rangle
\end{align}
where we have made 
the choice $x_{t_a} < y_a < x_{t_a}+1$ (which fixes the magnetic
field felt by the untwisted state) ,
we deduce
\begin{align}
&
\langle \sigma_{\epsilon_1, \kappa_1}(x_1, \bar x_1)\dots
\sigma_{\epsilon_N, \kappa_N}(x_N, \bar x_N)
\partial_{y_1} Z(y_1,\bar y_1) \partial_{y_2} {\bar Z}(y_2,\bar y_2)
\rangle
\nonumber\\
&=
\langle V_{N+2}(\{x_t\}_{t=1,\dots N}, \{x_a\}_{a=1.2}|
\otimes_t |T_t\rangle 
~\otimes (-i \sqrt{2\alpha'} \cos\gamma_{t_1}) \alpha_{(1) 1}^\dagger |0_{(1)}\rangle
~\otimes (-i \sqrt{2\alpha'} \cos\gamma_{t_2}~) \bar \alpha_{(2) 1}^\dagger |0_{(2)}\rangle
\nonumber\\
&=
\delta(\sum_t \kappa_t)
e^{-\oh \sum_{t} \kappa_t^2  ~G^{2 2}_{bou,~reg~T(t)}(x_t; x_t)}
e^{-\oh \sum_{u,t} \kappa_t \kappa_u ~G^{2 2}_{bou}(x_t; x_u)}
~C(x_1,\dots x_N) 
~\partial_{y_1} \partial_{y_2} 
\Gr^{z \bar z}_{bou}(y_1; y_2; \{x_t\})
\end{align}
where $\by$ is a function of $y$ as in eq. (\ref{barx-function-of-x}).
The previous result
is an immediate consequence of the the definition of $\Gr$ given in
(\ref{Green-full}) but it can also be interpreted as a ``proof'' of
eq. (\ref{Green-full}) since the Green function entering in the
previous formula is the Green function obtained from the path integral.

%If we now take the limit $x_{(1,t_1)} \rightarrow x_t$ we can use the
%OPE

\subsection{Example 3: $N-1$ not excited twisted, $1$ excited twisted
  and $2$ untwisted states}
We can now discuss the excited twisted states.
The easiest way to denote an excited twisted state is by writing from
which untwisted state it can be obtained by OPE, for example
$  \left(\partial_x Z ~\partial_x^2 Z ~\sigma_{\epsilon, \kappa}\right)(x,\bx)$ 
can be obtained by taking the finite part of the OPE
$(\partial_y Z ~\partial_y^2 Z)(y,\by) \sigma_{\epsilon, \kappa}(x,\bx) $ as
$y\rightarrow x$.
This limit can be taken in a clearer and easier way when
we realize $(\partial_y Z~\partial_y^2 Z)(y,\by)$ as a normal ordered operator in
a twisted auxiliary Hilbert space where $\sigma_{\epsilon, \kappa}(x,\bx) 
\leftrightarrow  |T, \kappa\rangle$.
In particular we can define the finite part of the limit as
\begin{equation}
\lim_{y\rightarrow 0^+} 
:
\left[
y^{1-\epsilon} \partial_y  Z^{}(y,y)
~\partial_y[ y^{1-\epsilon} \partial_y  Z^{}(y,y)]  
\right]
:
| T_{}, \kappa\rangle 
=
\left( -i \sqrt{2\alpha'} 
\,\cos\gamma
\right)^2
\alpha^{\dagger}_{ \epsilon}
\alpha^{\dagger}_{ 1+\epsilon}
 | T, \kappa\rangle 
\end{equation}

Similarly we can consider the map
\begin{align}
&
\left( \partial \bar Z \sigma_{\epsilon_t,\kappa_t}\right)(x_t,\bar x_t) 
\leftrightarrow
  -i \sqrt{2\alpha'} \,\cos\gamma_t \bar \alpha^{\dagger}_{1- \epsilon}
|T_t, \kappa_t\rangle
\end{align}
and compute the correlator
\begin{align}
&
\langle 
\left( \partial_x {\bar Z}\sigma_{\epsilon_1, \kappa_1}\right)(x_1, \bar x_1)\dots
\sigma_{\epsilon_N, \kappa_N}(x_N, \bar x_N)
\partial_{y_1} Z(y_1,\bar y_1) 
\rangle
\nonumber\\
&=
\langle V_{N+1}(\{x_t\}_{t=1,\dots N}, \{x_a\}_{a=1.2}|
(-i \sqrt{2\alpha'} \cos\gamma_{1}~) 
\bar \alpha_{(1)  \epsilon_1}^\dagger |T_1, \kappa_1\rangle
\otimes_{t>1} |T_t, \kappa_t\rangle 
~\otimes (-i \sqrt{2\alpha'} \cos\gamma_{t_1}) \alpha_{(1) 1}^\dagger |0_{(1)}\rangle
\end{align}
to be
\begin{align}
& \delta(\sum_t \kappa_t)
e^{-\oh \sum_{t} \kappa_t^2  ~G^{2 2}_{bou,~reg~T(t)}(x_t; x_t)}
e^{-\oh \sum_{u,t} \kappa_t \kappa_u ~G^{2 2}_{bou}(x_t; x_u)}
~
C(x_1,\dots x_N)~
\nonumber\\
&~~
\partial_{y_1}
\left[(x-x_1)^{\epsilon_1} 
\partial_x \Gr^{\bar z z}_{bou}(x, y_1; \{x_v\})
\right]|_{x=x_1}
.
\end{align}
The same computation can be performed using the generating function as
\begin{align}
&
\langle 
\left( \partial_x {\bar Z}\sigma_{\epsilon_1, \kappa_1}\right)(x_1, \bar x_1)\dots
\sigma_{\epsilon_N, \kappa_N}(x_N, \bar x_N)
\partial_{y_1} Z(y_1,\bar y_1) 
\rangle
=
\cV_{N+M} 
\frac{ \stackrel{\leftarrow}{\partial} }{ \partial d_{(1) 1} }
e^{i \kappa_1  
\frac{ \stackrel{\leftarrow}{\partial} }{\partial d_{(1) 0}^2 } }
\prod_{t>1} 
e^{i \kappa_t  
\frac{ \stackrel{\leftarrow}{\partial} }{\partial d_{(t) 0}^2 }
}
\frac{ \stackrel{\leftarrow}{\partial} }{\partial c_{(1) 1} }
\Big|_{c=0; d=0}
\end{align}

\section{Derivation for untwisted matter}
\label{sect:untwisted}
The starting point is very similar to
(\cite{DiVecchia:1986mb},\cite{Petersen:1988cf}) where it was
recognized that the generator for all closed (super)string amplitudes 
is a quadratic path integrals.
The idea in the previous papers is that the appropriate boundary
condition for R and/or NS sector can be obtained simply by inserting
{\sl linear} sources with the desired boundary conditions.
Because of this assumption the quantum fluctuations are the same for
all the amplitudes: from the purely NS to the mixed ones.
It was later realized that this prescription misses a proper treatment
of the quantum fluctuations (\cite{DiVecchia:1989hf}) 
and that when this part is considered the
amplitudes factorize correctly (\cite{DiBartolomeo:1990fw}).

Here we consider open strings and we realize the proper twisted
boundary conditions by {\sl quadratic} boundary terms which are
nothing else but the coupling of the string to the magnetic field background.
Therefore a non excited twist field is realized by a discontinuity in the
magnetic field\footnote{
And in particular the transition from an eigenstate $|\alpha\rangle$ in 
magnetic field $B_t$ 
to an eigenstate $|\beta\rangle$ in magnetic
$B_{t+1}$ at worldsheet time $\tau_t$ can be computed as in usual
quantum mechanics as 
$\langle \beta | \alpha \rangle =
\int \cD X ~ 
\langle X(\sigma), \tau_t^- | \alpha \rangle
\left( \langle X(\sigma), \tau_t^+| \beta \rangle\right)^*$.
}.

The $N$ non excited twist field amplitudes with Euclidean worldsheet
metric is then computed by the quadratic path integral\footnote{
We define $d^2 z= 2 d z ~ d \bar z$ and $z=e^{\tau_E+i \sigma}\in H$.
}
\footnote{This is the path integral corresponding to all twist fields
  with zero ``momentum'' and therefore it is proportional to a
  $\delta(0)$  which arises from $X^2$ zero mode. The general case is
  treated in the next section.}
\COMMENTOOK{$X(x,x)\ne X(x,\bar x)$ when $x<0$. Is this way of writing
correct also in this case? On the other side we can start with all
$x>0$ by using a conformal transformation so the case where some $x<0$
must be reached from the case where this writing is right.
Almost since the simple case with $N=2$ reveals that it is important
to keep $x\ne \bar x$ since in this way we see that the divergence on
the two boundaries does depend on the magnetic field felt on the
boundary of interest.
In any case $d x  \partial_x X^2(x,\bx)$ must be interpreted as the
pullback on the boundary of $d X^2$ in such a way that $\bar x$
depends on $x$.}
\begin{align}
\label{Z0}
C(x_1,\dots x_N)
\delta(0)
&=
\cN
\int \cD X~
e^{
-\frac{1}{ 2\pi \alpha'}\left[
\int_H d^2z ~ \oh G_{i j} \partial_z X^i \partial_{\bar z} X^j
-i \int_{\partial H} d x~ B(x) X^1(x,\bx) \partial_x X^2(x,\bx)
\right]
} 
\end{align}
where $d x  \partial_x X^2(x,\bx)$ must be interpreted as the
pullback on the boundary of $d X^2$ in such a way that $\bar x$
depends on $x$ as
\begin{equation}
\label{barx-function-of-x}
\bx = \left\{\begin{array}{ll}
x & x=|x|>0
\\
x e^{-i 2\pi} & x=|x| e^{i\pi}<0
\end{array}
\right.
,
\end{equation}
$H$ is the superior half plane and the adimensional magnetic
field $B(x)$ is given by  
\begin{equation}
B(x)
= 2\pi \alpha' ~q F_{1 2}(x)
= \sum_{t=1}^N B_t ~\theta(x-x_t) ~\theta(x_{t+1}-x)
\end{equation}
so that the dicharged string at $x=x_t$ feels a magnetic field
$B_{t-1}$ on the left, i.e. $\sigma=\pi$  and
$B_{t}$ on the right, i.e. $\sigma=0$.
Here we have set $B_{-1}=B_{N+1}$ and $x_{-1}=-\infty$, $x_{N+1}=+\infty$.
\COMMENTOOK{ cases $t=0$ and $t=N+1$ connection with $\epsilon$}

In the previous equation (\ref{Z0}) we have chosen the gauge
\begin{equation}
A_1=0,
~~~~
A_2=B x^1
\end{equation}
in order to make clear that we have only one zero mode which is
associated with the shift $X^2 \rightarrow X^2 + \epsilon$ and
therefore there is only one conserved momentum as it is the case in
the Landau levels problem on $\R^2$.
\COMMENTOO{This is OK but what is the difference with the case with a
  Kalb-Ramond bck written as a magnetic field? In this case we have
  two zero modes and actually the eom with the bc are invariant under
  two shifts.
}

Now we can add the untwisted states vertexes.
This can be done by considering the generating function of all
untwisted vertexes at $x=x_a\in\R$ as\footnote{
This formulation involving the full field on the boundary is only
right for NN boundary condition. For DD boundary condition one should
use a slightly different one (\cite{Pesando:2003ww}).
}
%%%%%%%%%%
\COMMENTOOK{Again $X(x,x)$ or $X(x,\bar x)$?}
%%%%%%%%%%
\begin{align}
\cS(c_{(a)})
&= 
\exp\{
\sum_{n=0}^\infty c_{(a)\, n\, i} \partial^n_x|_{x=x_a} X^i(x,\bx)
\}
=
\exp\{ \int_{\partial H} d x ~ J_i(x;x_a) ~X^i(x,\bx)\}
\nonumber\\
J_i(x; x_a)
&=
\sum_{n=0}^\infty c_{(a) n i} \partial^n_{x_a}  \delta(x-x_a)
\end{align}
where $c$ are arbitrary complex numbers (or functions as we use in the
next section).
To understand how the previous generating vertex work 
let us take the example from (\cite{Pesando:2011yd}).
Consider the vertex which describes the fluctuations of the gauge vector
around the dipole string background 
we can derive it from a generating functional for the dipole string as
%%%%%%%%%%%%
\footnote{
 Notice that here we are talking about the abstract path integral
 representation of the vertex  and not of the operatorial representation.
The operatorial representation of the vertex
can be realized in an auxiliary space with both twisted and untwisted
boundary conditions. The untwisted auxiliary Hilbert space
representation is the usual operatorial representation while the
twisted one is the one derived in (\cite{Pesando:2011yd}).
Just because of these different realizations this auxiliary Hilbert
space must not be confused with $\cH_{(a,t_a)}$ introduced before
which is a way of representing the $c_{(a) n i}$, see (\ref{c-realization}).
}
%%%%%%%%%%
%%%%%%%%%%
%\COMMENTOO{
\footnote{
It is worth noticing how vertexes for dipole strings have
  the same functional form independently on the magnetic backgrounds
$B_{t_a}$
nevertheless they differ because different conditions for physical
states, in the previous example we have 
%$ 
%k_{(a,t_a) \mu} \eta^{\mu \nu} k_{(a,t_a)\nu} 
%+ k_{(a, t_a) i} \cG^{i j}(B_{t_a})  k_{(a,t_a) j} = 
%\epsilon_{(a,t_a) \mu} \eta^{\mu \nu} k_{(a,t_a) \nu}
%+ \epsilon_{(a,t_a) i} \cG^{i j}(B_{t_a})
%k_{(a, t_a) j}
%=0
%$
$ 
k_{\mu} \eta^{\mu \nu} k_{\nu} 
+ k_{i} \cG^{i j}(B_{t_a})  k_{j} = 
\epsilon_{\mu} \eta^{\mu \nu} k_{\nu}
+ \epsilon_{ i} \cG^{i j}(B_{t_a})
k_{ j}
=0
$
where $\mu,\nu\ne 1,2$.
}
%%%%%%%%%%%
\begin{align}
V(x_a;\epsilon,k)=
\epsilon_i \partial_x X^i(x,\bx) e^{ i k_j X^j(x,\bx)}
\Big|_{x=x_a}
=
S(c_{(a)},x_a) ~\epsilon_i 
\frac{ \stackrel{\leftarrow}{\partial} }{ \partial c_{(a) 1 i} } 
e^{ i k_j \frac{    \stackrel{\leftarrow}{\partial} }{\partial c_{(a) 0 j} } }
\Big|_{c_{(a)}=0}
\end{align}

As a matter of facts the previous vertex gives an indefinite result
when inserted in the path integral even when $B=0$.
We must therefore regularize it and consider
\begin{align}
\label{SDS}
\cS_{reg}(c_{(a)})
&= 
\cN_{(a)}(x_a)
\exp\Big\{
\sum_{n=0}^\infty c_{(a)\, n\, i} \partial^n_x|_{x=x_a} \langle X^i (x,\bx)\rangle
\Big\}
\nonumber\\
&=
\cN_{(a)}(x_a)
\exp\Big\{ 
\int_{\partial H} d x ~ J_{i, reg}(x;x_a) ~X^i(x,\bx)
\Big\}
\end{align}
where the regularized curred is given by
\begin{align}
J_{i, reg}(x; x_a)
&=
\sum_{n=0}^\infty c_{(a) n i} \partial^n_{x_a}  \delta_{reg}(x-x_a)
,
\end{align}
the averaged field by
\begin{equation}
\langle X^i (x,x)\rangle= \int_{\partial H} d y~ \delta_{reg}(x-y)
X(y,y)
\end{equation}
and the normalization factor is
%%%%%%%%%%
\COMMENTOOK{There is probably a factor $2$ due to the boundary
  regularization.
No there is not since we start with the open Green function and not
with the closed one.
}
%%%%%%%%%%
\begin{align}
\cN_{(a)}(x_a)
= 
\exp\Big\{
&
-
\oh \sum_{n,m=0}^\infty 
c_{(a) n i} ~c_{(a) m j}
\nonumber\\
&
\int_{\partial H} d x \int_{\partial H} d y
~\partial^n_x%|_{x=x_a} 
\delta_{reg}(x-x_a)
~\partial^m_y%|_{y=x_a} 
\delta_{reg}(y-x_a)
~\Gr_{U(t_a),~ bou}^{i j}(x; y)
\Big\}
\Big|_{x=x_a,y=x_a} 
\end{align}
with $\Gr_{U(t_a),~bou}^{i j}(x; y ) $ the boundary 
Green functions in the dipole case
with magnetic field $B_{t_a}$  given in eq.s (\ref{dipole-Gr-bou}).
There are two reasons why we have introduced the previous definitions.
The first is that it works in reproducing the amplitudes for $N=2$ as
discussed in appendix \ref{app:N=2-check}.
The second is connected to the way the regularization terms is
suggested from the operatorial formalism. In operatorial formalism the
simplest approach is to consider a point splitting, i.e.
$[\exp \left( c_{(a) i} X^i_{(a)}(x_a,x_a)\right)]_{p.s.}
=
\exp \left( c_{(a) i} [ X^{i(-)}_{(a)}(x_a e^{-\eta}, x_a e^{-\eta})
+
X^{i (+)}_{(a)}(x_a, x_a)]
\right)
$ which implies a regularization factor
$\cN_{(a)}(x_a)= \exp\left(
-\oh c_{(a) i} c_{(a) j} G^{i j}_{bou}(x; y)
\right)\Big|_{x=x_a, y=x_a e^{-\eta}}
$.
When we smooth the fields the previous regularization factor becomes
\begin{align}
\cN_{(a)}(x_a)
= &
\exp\left(
-\oh c_{(a) i} c_{(a) j} \int_{x>y} d x ~d y~ 2~\delta_{reg}(x-x_a)
%~\partial^m_y|_{y=x_a}
\delta_{reg}(y-x_a) G^{i j}_{bou}(x; y)
\right)%\Big|_{x=x_a, y=x_a e^{-\eta}}
\nonumber\\
=&
\exp\left(
-c_{(a) i} c_{(a) j} \int d x ~d y~ \delta_{reg}(x-x_a)
%~\partial^m_y|_{y=x_a}
\delta_{reg}(y-x_a) 
\frac{ 
G^{i j}_{bou}(x; y)~ \theta(x-y)
+
G^{j i}_{bou}(y; x)~ \theta(y-x)
}{2}
\right)%\Big|_{x=x_a, y=x_a e^{-\eta}}
\nonumber\\
=&
\exp\left(
-\oh c_{(a) i} c_{(a) j} \int d x ~d y~ \delta_{reg}(x-x_a)
%~\partial^m_y|_{y=x_a}
\delta_{reg}(y-x_a) 
G^{i j}_{bou}(x; y)
\right)%\Big|_{x=x_a, y=x_a e^{-\eta}}
\end{align}
where the factor $2$ in the first line is due to the fact we are using
one of the two $\delta$ just one half because of the constraint $x>y$
as it can be directly verified by using a step function regularization
of the delta.
In the last step we have used the property $G^{j i}_{bou}(y; x) = G^{i j}_{bou}(x; y) $.

In conclusion the path integral we want to compute in order to get the
generating function for all the $M$ untwisted correlators in presence
of $N$ twists is 
%%%%%%%%%%%
\COMMENTOO{
It is not necessary to set
$2\pi \alpha' =1$
since the Green function obtained as $[X^{(+)}(x,\bx), X^{(-)}(y, \by)]
$
are correctly normalized as it can be verified for the the dipole (and
hence neutral) strings.
}
%%%%%%%%%%%
\begin{align}
Z(\{x_t\}; \{x_a\})
&=
\cN
\int \cD X~
e^{
-\frac{1}{2\pi \alpha'} \left[
\int_H d^2 z ~ \oh G_{i j} \partial_z X^i \partial_{\bar z} X^j
+i \int_{\partial H} d x~ B(x) X^1(x,\bx) \partial_x X^{2}(x,\bx)
\right]
}
\nonumber\\
&~~~~\times
\prod_{a=1}^M \cN_{(a)}(x_a)
e^{\int_{\partial H} d x~ J_{i, reg}(x; x_a) X^i(x,\bx) }
\end{align}

This path integral can then be performed to get
%%%%%%%%%%%
\COMMENTOO{
1) zero mode effect, probably only $\delta(\sum_a c_{(a) 0 \,i=2})$ \\
2) non commutative phase from Green function ? \\
3) non hermiticity of the quadratic operator implies eigenfunctions
are not orthogonal
}
%%%%%%%%%%
\begin{align}
\label{PI_untwisted_0}
Z(\{x_t\}; \{x_a\})
&=
C(x_1,\dots x_N)~\delta(i \sum_a c_{(a) 0 \,i=2})
\nonumber\\
&
\prod_a
\exp\Big\{
\oh 
\int_{\partial H} d x \int_{\partial H} d y
J_{i, reg}(x;x_a)~J_{j, reg}(y;x_a)
\nonumber\\
&
\hspace{5em}
~[\Gr^{i j}_{bou}(x; y;\{x_v\})-\Gr_{U(t_a)~,bou}^{i j}(x ; y )]
\Big\}
\nonumber\\
&
\prod_{a,b; a\ne b}
\exp\Big\{
\oh 
\int_{\partial H} d x \int_{\partial H} d y
J_{i, reg}(x;x_a)~J_{j, reg}(y;x_b)
\nonumber\\
&
\hspace{5em}
~\Gr^{i j}_{bou}(x ; y ; \{x_v\})
\Big\}
\end{align}
where $\Gr^{i j}_{bou}(x ; y; \{x_v\})$ is the Green function of the
quadratic operator which turns out to be the one defined in
(\ref{Green}).
Notice that the quadratic operator is not even hermitian exactly as it
happens for Landau levels in the plain quantum mechanics.
\COMMENTOO{
1) 
To justify this we can consider 
 the analogous path integral (\ref{Z0})  in the $(\tau, \sigma)$ variables
and then  discretize it in the $\sigma$ direction in such a way there
is no difference between worldsheet bulk and boundary terms...\\
 2)
Can we derive 
conformal invariance and on-shell conditions which depend on the
segment of $\partial H$? Not really, even in the neutral case  we can
only see that the dependence on the conformal factor which appears
through the form of the Green function decouples when momentum is
imposed.
Here we have only one zero mode and only one conservation. 
}
The previous result
can be also rewritten after the regularization has been removed as
\begin{align}
\label{PI_untwisted_1}
Z(\{x_t\}; \{x_a\})
=&
C(x_1,\dots x_N)~\delta(i \sum_a c_{(a) 0 \,i=2})
\nonumber\\
&
\prod_a
\exp\Big\{
\oh \sum_{n,m=0}^\infty c_{(a) n i} ~c_{(a) m j}
%\int_{\partial H} d x \int_{\partial H} d y
~\partial^n_x|_{x=x_a} %\delta_{reg}(x-x_a)
~\partial^m_y|_{y=x_a} %\delta_{reg}(y-x_a)
~\Gr^{i j}_{bou~reg~U(t_a)}(x; y; \{x_v\})
\Big\}
\nonumber\\
&
\prod_{a,b; a\ne b}
\exp\Big\{
\oh \sum_{n,m=0}^\infty c_{(a) n i} ~c_{(b) m j}
%\int_{\partial H} d x \int_{\partial H} d y
~\partial^n_x|_{x=x_a} %\delta_{reg}(x-x_a)
~\partial^m_y|_{y=x_b} %\delta_{reg}(y-x_a)
~\Gr^{i j}_{bou}(x ; y; \{x_v\})
\Big\}
\end{align}
where we have defined
the boundary Green function regularized by the untwisted Green function
for a background $B_{t_a}$
\begin{align}
\label{Green-bou-reg-U}
\Gr^{z \bar z}_{bou~reg~U (t_a)}(x, y; \{x_v\})
&=
%\oh
\left[
\Gr^{z \bar z}_{bou}(x; y; \{x_v\})
-
\Gr^{z \bar z}_{U (t_a),~bou }(x; y)
\right]
%\nonumber\\
%&+
%\oh\left[
%\Gr^{z \bar z}_{bou}(y; x; \{x_v\})
%-
%\Gr^{z \bar z }_{U (t_a),~bou }(y; x)
%\right]
\end{align}

The previous expression can be simplified a little using the symmetry 
$G_{bou}^{i j}(x;y)=G_{bou}^{j  i}(y; x)$ and 
by rewriting it in the complex basis as
\begin{align}
\label{PI_untwisted_1z}
Z(\{x_t\}; \{x_a\})
=&
C(x_1,\dots x_N)~\delta\left(\sum_a (c_{(a) 0}- \bar c_{(a) 0}) \right)
\nonumber\\
&
\prod_a
\exp\Big\{
\oh
c_{(a) 0}^2 ~\Gr^{\bar z \bar z}_{bou~reg~U(t_a)}(x; y; \{x_v\})
+
\oh
\bar c_{(a) 0}^2 ~\Gr^{z z}_{bou~reg~U(t_a)}(x; y; \{x_v\})
%\Big\}
\nonumber\\
&
\phantom{  \prod_a \exp\Big\{ }
\sum_{n,m=0}^\infty \bar c_{(a) n} ~ c_{(a) m}
%\int_{\partial H} d x \int_{\partial H} d y
~\partial^n_x|_{x=x_a} %\delta_{reg}(x-x_a)
~\partial^m_y|_{y=x_a} %\delta_{reg}(y-x_a)
~\Gr^{z \bar z}_{bou~reg~U(t_a)}(x; y; \{x_v\})
\Big\}
\nonumber\\
&
\prod_{ a < b}
\exp\Big\{
\bar c_{(a) n} ~ \bar c_{(b) m}
~\Gr^{z z}_{bou}(x_a ; x_b; \{x_v\})
+
 c_{(a) n} ~ c_{(b) m}
~\Gr^{\bar z \bar z}_{bou}(x_a ; x_b; \{x_v\})
%\Big\}
\nonumber\\
&
\phantom{ \prod_{ a < b} \exp\Big\{ }
+ \sum_{n,m=0}^\infty \bar c_{(a) n} ~ c_{(b) m}
%\int_{\partial H} d x \int_{\partial H} d y
~\partial^n_x|_{x=x_a} %\delta_{reg}(x-x_a)
~\partial^m_y|_{y=x_b} %\delta_{reg}(y-x_a)
~\Gr^{z \bar z}_{bou}(x ; y; \{x_v\})
\Big\}
\nonumber\\
&
\phantom{ \prod_{ a < b} \exp\Big\{ }
+ \sum_{n,m=0}^\infty c_{(a) n} ~ \bar c_{(b) m}
%\int_{\partial H} d x \int_{\partial H} d y
~\partial^n_x|_{x=x_a} %\delta_{reg}(x-x_a)
~\partial^m_y|_{y=x_b} %\delta_{reg}(y-x_a)
~\Gr^{\bar z z}_{bou}(x ; y; \{x_v\})
\Big\}
\end{align}
with $\bar c_{(a) n} = c_{(a) n z}$ and $c_{(a) n} = c_{(a) n \bar z}$.

We can now give a different formulation of the previous result if we
realize the algebra 
\begin{equation} 
[ c_{(a) n  i}, 
\frac{ \stackrel{\leftarrow}{\partial} }{ \partial  c_{(b) m j}}]
= \delta_i^j~ \delta_{n, m}~ \delta_{a, b}
\end{equation}
on the untwisted scalar (dipole string) auxiliary Hilbert spaces $\cH_{a, t_a}$
with backgrounds $B_{t_a}$ introduced before as
\begin{align}
\label{c-realization}
1 
\rightarrow &
\langle z_{(a) 0 0}=\bar z_{(a) 0 0}=0| \langle 0_{(a)} |
\nonumber\\
\bar c_{(a) n\,} 
\rightarrow \frac{i}{\sqrt{2\alpha'}  \cos \gamma_{t_a}}
\frac{\alpha_{(a) n}}{n!}
,&~~
c_{(a) n\, } 
\rightarrow \frac{i}{\sqrt{2\alpha'}  \cos \gamma_{t_a}}
\frac{\bar \alpha_{(a) n}}{n!}
~~~ n\ge 0
\nonumber\\
\frac{ \stackrel{\leftarrow}{\partial} }{ \partial  \bar c_{(a) m}}
\rightarrow
-i \sqrt{2 \alpha'}\, (m-1)!\, \cos \gamma_{t_a} \, \alpha_{(a) m}^\dagger
,&~~
\frac{ \stackrel{\leftarrow}{\partial} }{ \partial  c_{(a) m }}
\rightarrow
-i \sqrt{2 \alpha'}\, (m-1)!\, \cos \gamma_{t_a}\, \bar \alpha_{(a) m}^\dagger
~~ m>0
\end{align}
where the ``strange'' choice of the normalization 
is due to the last expressions which arise from the desire of
identifying
\begin{equation}
-i \sqrt{2 \alpha'}\, (m-1)!\, \cos \gamma_{t_a}\, \alpha_{(a) m}^\dagger
\sim 
\partial^m X^{(-)z}_{(a)}(x, \bx) |_{x=0}
.
\end{equation}
Using these auxiliary Hilbert spaces we can now rewrite the previous
expression for the $M$ untwisted correlators (\ref{PI_untwisted_1z}) as 
\begin{align}
\label{PI_untwisted_op_1_0}
Z(\{x_t\}; \{x_a\})
&
=
C(x_1,\dots x_N)~
\delta\left(i \sum_a (\alpha_{(a) 0 } -\bar \alpha_{(a) 0 })\right)
~\prod_{a=1}^M \langle z_{(a) 0 0}=\bar z_{(a) 0 0}=0| \langle 0_{(a)} |
\nonumber\\
&
\prod_a
\exp\Big\{
-\frac{1}{4\alpha'}
\alpha_{(a) 0}^2 
~\cV_{(t_a)~\underline{ z} \bar z}^2
~\Gr^{\bar z \bar z}_{bou,~reg~U(t_a)}(x; y; \{x_v\})
\nonumber\\
&
%%%%%%%%%%
\phantom{\prod_a\exp\Big\{}
%%%%%%%%%%
-
\frac{1}{4\alpha'}
\bar \alpha_{(a) 0}^2 
~\cV_{(t_a)~\underline{\bar z} z}^2
~\Gr^{z z}_{bou,~reg~U(t_a)}(x; y; \{x_v\})
\nonumber\\
&
%%%%%%%%%%
\phantom{\prod_a\exp\Big\{}
%%%%%%%%%%
-\frac{1}{2\alpha'}
 \sum_{n,m=0}^\infty \alpha_{(a) n} ~\bar \alpha_{(a) m}
~\cV_{(t_a)~\underline{\bar z} z}\cV_{(t_a)~\underline{ z} \bar z }
~\frac{ \partial^n_x }{ n! } 
~\frac{ \partial^m_y }{ m!}
~\Gr^{z \bar z}_{bou,~reg~U(t_a)}(x; y; \{x_v\})
\Big\}
\Big|_{x=y=x_a}
\nonumber\\
&
\prod_{a<b}
\exp\Big\{
-\frac{1}{2\alpha'}
\sum_{n,m=0}^\infty \alpha_{(a) n } ~\bar \alpha_{(b) m}
~\cV_{(t_a)~\underline{\bar z} z}\cV_{(t_b)~\underline{ z} \bar z }
~\frac{ \partial^n_x }{ n! } 
~\frac{ \partial^m_y }{ m!}
~\Gr^{z \bar z}_{bou}(x ; y; \{x_v\})
%\Big\}
\nonumber\\
&
%%%%%%%%%
\phantom{\prod_{a<b}\exp\Big\{ }
%%%%%%%%%
-\frac{1}{2\alpha'}
\sum_{n,m=0}^\infty \bar \alpha_{(a) n } ~\alpha_{(b) m}
~\cV_{(t_b)~\underline{\bar z} z}\cV_{(t_a)~\underline{ z} \bar z }
~\frac{ \partial^n_x }{ n! } 
~\frac{ \partial^m_y }{ m!}
~\Gr^{\bar z z}_{bou}(x ; y; \{x_v\})
\Big\}
\Big|_{x=x_a, y=x_b}
\end{align}
where $\cV$ are the $\R^2$ vielbein which connect the $\alpha$ flat
index with the Green function $G$ curved index.

\section{Derivation for twisted matter}
\label{sect:twisted}
The strategy we are going to follow is to consider the amplitude
derived in previous section with
$N+M$ untwisted states  at the positions 
$\{x_a\}_{a=1\dots M}, \{x_f\}_{f=1\dots N}$
and unexcited twists at positions $\{x_t\}_{t=1\dots N}$.
Then we choose $N$ of untwisted states at the positions  $ \{x_f\}_{f=1\dots N}$
for which we take
the limit $x_f \rightarrow x_t$.
In order to get the desired amplitude with $M$ untwisted and $N$
excited twisted states we must
choose in a proper way the $c_{(f) n i}$.
This amounts not only to choose  $c_{(f) n i}$ 
 in (\ref{SDS}) as a function of $x_f$  as in eq. (\ref{c(d,x)})
but to introduce a further
 normalization $\cR(x_f)$ as in eq. (\ref{R(d,x)})
in such a way that we can ``undo'' the OPE and get
a result which is a generating function for the twisted states 
\begin{equation}
\label{coherent-twist-state}
\exp\Big\{
d^2_{(t)0} x_0^{i=2 (aux\,t)}
-
i \sqrt{2\alpha'} 
\,\cos\gamma_t
\,
\sum_{n=1}^\infty \left[
d_{(t) n} 
%\{(n-1)!}{\sqrt{n+(1-\epsilon_t)}} 
(n-1)!
\bar \alpha^{\dagger (aux\,t)}_{ n-\epsilon_t} 
+
\bar d_{(t_f) n}  
%{(n-1)!}{\sqrt{n+\epsilon_t}} 
(n-1)!
\alpha^{\dagger (aux\,t)}_{ n-1+\epsilon_t} 
\right]
\Big\}
| T_{(aux\,t)}\rangle
\end{equation}
when expressed in a chart where $x_t=0$. The ``strange'' normalization
is chosen because it is the easiest map from operators to states, f.x.
the twisted excited state which can be obtained by subtracting the
divergences of the limit
$y\rightarrow x_t^+$ of $[\partial_y^3 Z(y,y)]^2 \sigma_{\epsilon_t}(x_t,x_t)$
gives the state
\begin{equation}
\lim_{y\rightarrow 0^+} 
\left[
\partial_y^2[ y^{1-\epsilon_t} \partial_y  Z^{(aux\,t)}(y,y)]  
\right]^2
| T_{(aux\,t)}\rangle 
=
\left( -i \sqrt{2\alpha'} 
~\cos\gamma_t ~2! ~\alpha^{\dagger (aux\,t)}_{ 2+\epsilon_t}
\right)^2
 | T_{(aux\,t)}\rangle 
\end{equation}
thus making contact between eq. (\ref{coherent-twist-state}) and
eq. (\ref{SDS-x=0}).

Let us start studying the OPE $\cS(c, x_f) \sigma_{t}(0,0)$.
This can be studied in an auxiliary Hilbert space $\cH_{aux\, t}$
(not to be confused the the Hilbert space $\cH_{t}$ which we
introduced in the first section 
and which is associated with coefficients $d_{(t)}$)
where $ \sigma_{\epsilon_t}(0,0)$ is represented by the twisted vacuum 
$|T_{(aux\,t)}\rangle$  and the generating function $\cS(c, x_f)$ as
\begin{align}
 \cS_{(aux\,t)}(c_{(f)}, x_f)
=&
e^{ 
-\oh
c_{(f) 0 }^2 %~c_{(f) 0 }
~\Gr^{z z}_{bou,~reg~U(t_a)}(x_f; x_f; \{x=0, x=\infty\})
-\oh
\bar c_{(f) 0 }^2 %~\bar c_{(f) 0 }
~\Gr^{\bar z \bar z}_{bou,~reg~U(t_a)}(x_f; x_f; \{x=0, x=\infty\})
}
\nonumber\\
&
e^{ 
-\oh
\sum_{m,n=0}^\infty
c_{(f) n } ~\bar c_{(f) m }
~\partial^n_x|_{x=x_f}
~\partial^m_y|_{y=x_f} 
~\Gr^{z \bar z}_{bou,~reg~U(t_a)}(x; y; \{x=0, x=\infty\})
}
\nonumber\\
&
:
e^{
\sum_{n=0}^\infty
\bar c_{(f) n }  ~\partial^n_x|_{x=x_f} Z_{(aux\,t)}(x, \bar x)
+
c_{(f) n }  ~\partial^n_x|_{x=x_f} \bar Z_{(aux\,t)}(x, \bar x)
}
:
\end{align}
In the previous equation the normal ordering is performed with
respect to the operators entering the expansion of the quantum fields
$Z_{(aux\,t)}(z, \bar z)$ and $\bar Z_{(aux\,t)}(z, \bar z)$ which act on
$\cH_{aux\, t}$.
Then the OPE can be computed as
\begin{align}
\cS(c, x_f) \sigma_{\epsilon_t}(0,0)
\leftrightarrow
&
e^{ 
-\oh
c_{(f) 0 }^2 %~c_{(f) 0 }
~\Gr^{z z}_{bou,~reg~U(t_a)}(x_f; x_f; \{x=0, x=\infty\})
-\oh
\bar c_{(f) 0 }^2% ~\bar c_{(f) 0 }
~\Gr^{\bar z \bar z}_{bou,~reg~U(t_a)}(x_f; x_f; \{x=0, x=\infty\})
}
\nonumber\\
&
e^{ 
-\oh
\sum_{m,n=0}^\infty
\partial^n_x|_{x=x_f}
~\partial^m_y|_{y=x_f} 
\left[
~c_{(f) n } ~\bar c_{(f) m }
~\Gr^{z \bar z}_{bou,~reg~U(t_a)}(x; y; \{x=0, x=\infty\})
\right]
}
\nonumber\\
&
e^{
\sum_{n=0}^\infty
\partial^n_x|_{x=x_f} \left[ \bar c_{(f) n } ~Z_{(aux\,t)}^{(-)}(x,x)\right]
+
\partial^n_x|_{x=x_f} \left[ c_{(f) n }  ~\bar Z_{(aux\,t)}^{(-)}(x,x)\right]
}
| T_{(aux\,t)}\rangle
\end{align}
which is similar to a rewriting of eq. (\ref{coherent-twist-state}) as
\begin{align}
\label{SDS-x=0}
\lim_{x_f\rightarrow 0^+}
&
e^{
\bar d_{(t) 0 }  ~\left[  Z_{(aux~t)}^{(-)}(x_f, x_f) \right]
+
d_{(t) 0}  ~\left[ \bar   Z_{(aux~t)}^{(-)}(x_f, x_f) \right]
}
\nonumber\\
&
e^{
\sum_{n=1}^\infty
\bar d_{(t) n }  
~\partial^{n-1}_x|_{x=x_f} \left[ x^{1-\epsilon_t} 
\partial_x Z_{(aux~)}^{(-)}(x,x) \right]
+
d_{(t) n }  
~\partial^{n-1}_x|_{x=x_f} \left[ x^{\epsilon_t} 
\partial_x \bar  Z_{(aux~)}^{(-)}(x,x) \right]
}
| T_{(aux\,t)}\rangle
\end{align}
where in the second line we have written $\partial Z$ since we want to
get rid of zero modes
and
in the limit it is necessary to write $x_f\rightarrow 0^+$ since the
the behavior of $\partial Z$ changes by an overall normalization when $x<0$.

Comparison between the two previous expressions 
suggests to consider then the operator acting on the
Hilbert space $\cH_{(aux~t)}$
\begin{align}
\label{op-T}
\cT_{(aux~t)}(d_{(t)},x_f)
&=
\cN(d_{(t)},x_f,x_t) 
~
e^{
\bar d_{(t) 0 }  ~\left[  Z_{(aux~t, reg)}(x_f, x_f) \right]
+
d_{(t) 0}  ~\left[ \bar   Z_{(aux~t, reg)}(x_f, x_f) \right]
}
\nonumber\\
&
e^{
\sum_{n=1}^\infty
\bar d_{(t) n }  
~\partial^{n-1}_x \left[ x^{1-\epsilon_t} 
\partial_x Z_{(aux~t, reg)}(x,x) \right]
+
d_{(t) n }  
~\partial^{n-1}_x \left[ x^{\epsilon_t} 
\partial_x \bar  Z_{(aux~t, reg)}(x,x) \right]
}
\Big|_{x=x_f^+}
\end{align}
where $Z_{(aux~t, reg)}(x_f, x_f) $ is point split regularized of
$Z_{(aux~t)}(x_f, x_f) $ defined as
$$
Z_{(aux~t, reg)}(x_f, x_f) = 
Z_{(aux~t)}^{(-)}(x_f e^{-\eta}, x_f e^{-\eta}) +
Z_{(aux~t)}^{(+)}(x_f, x_f) , 
$$
no normal ordering is performed and the normalization factor is given
by
\begin{align}
\cN^{-1}(d_{(t)},x_f,x_t)
&=
\Big\{
e^{
\oh \bar d_{(t) 0 }^2 G^{z z}_{T(t)~bou}(x; y)
+
\oh  d_{(t) 0 }^2 G^{\bar z \bar z}_{T(t)~bou}(x; y)
+
 \bar d_{(t) 0 } d_{(t) 0 }  
\frac{G^{z \bar z}_{T(t)~bou}(x; y) +G^{\bar z z}_{T(t)~bou}(x; y) }{2}
}
\nonumber\\
&
e^{
\oh
\sum_{n=1}^\infty
\bar d_{(t) 0 } d_{(t) n}
\big[
\partial^{n-1}_x( x^{\epsilon_t} \partial_x  G^{\bar z z}_{T(t)~bou}(x; y))
+
\partial^{n-1}_y( y^{\epsilon_t} \partial_y  G^{z \bar z}_{T(t)~bou}(x; y))
\big]
}
\nonumber\\
&
e^{
\oh
\sum_{n=1}^\infty
d_{(t) 0 } \bar d_{(t) n}
\big [
\partial^{n-1}_x( x^{1-\epsilon_t} \partial_x  G^{z \bar z}_{T(t)~bou}(x; y))
+
\partial^{n-1}_y( y^{1-\epsilon_t} \partial_y  G^{\bar z z}_{T(t)~bou}(y; x))
\big]
}
\nonumber\\
&
e^{
\oh
\sum_{n,l=1}^\infty
d_{(t) l } \bar d_{(t) n}
\big [
\partial^{l-1}_y
\partial^{n-1}_x( 
y^{\epsilon_t} x^{1-\epsilon_t} \partial_x \partial_y  G^{z \bar  z}_{T(t)~bou}(x; y)
)
+
\partial^{l-1}_x
\partial^{n-1}_y( 
x^{\epsilon_t} y^{1-\epsilon_t} \partial_x \partial_y  G^{\bar z z}_{T(t)~bou}(x; y))
\big]
}
\nonumber\\
&\Big\}
\Big|_{x=x_f; y=x_f e^{-\eta}}
\end{align}
where $G^{i j}_{T(t)~bou}(y; x) = G^{j i}_{T(t)~bou}(x; y)=
G^{i j}_{bou}(y; x; \{x_1=0, x_2=\infty\})$ are the (analytic
continuation of the) boundary
Green functions defined in eq.s (\ref{twisted-bou-green}).

The reason why we have written the previous expression in a non normal
ordered way is to understand the expression of the regularization
factor of the corresponding ``classical'' vertex (\ref{vertex-T})
which we want to insert in the path integral.  

The previous operator can also be written in a way to make its
connection with the idea of undoing the OPE clearer as
\begin{align}
\cT_{(aux~t)}(d_{(t)},x_f)
=&
\cR(d_{(t)},x_f) ~\cS_{(aux~t)}(c_{(f)}(d_{(t)},x_f), x_f)
\nonumber\\
=&
\cR(d_{(t)},x_f)
e^{
\sum_{n=0}^\infty
\bar d_{(t) n }  
~\partial^n_x|_{x=x_f} \left[ x^{-\epsilon_t} Z_{(aux\,t)}(x,x) \right]
+
d_{(f) n }  
~\partial^n_x|_{x=x_f} \left[ x^{-(1-\epsilon_t)} \bar
  Z_{(aux\,t)}(x,x) \right]
}
\end{align}
where
\begin{align}
\label{c(d,x)}
c_{(f) 0}(d_{(t)},x_f) = d_{(t) 0}
~~&~~
\bar c_{(f) 0}(d_{(t)},x_f) =\bar d_{(t) 0}
\nonumber\\
c_{(f) n}(d_{(t)},x_f)
=&
\sum_{k=n}^\infty  \binom{k-1}{n-1} d_{(t) k} \partial^{k-n}
x_f^{-(1-\epsilon_t)}
\nonumber\\
\bar c_{(f) n}(d_{(t)},x_f)
=&
\sum_{k=n}^\infty  \binom{k-1}{n-1} \bar d_{(t) k} \partial^{k-n}
x_f^{\epsilon_t}
\end{align}
and the normalization factor is
\COMMENTOO{ Check!}
\begin{align}
\label{R(d,x)}
\cR(d_{(t)},x_f)
=&
e^{ 
-
\sum_{m,n=0}^\infty
d_{(t) n } ~\bar d_{(t) m }
~\partial^{n-1}_x
~\partial^{m-1}_y%|_{y = x_f}
\left[  x^{1-\epsilon_t}  y^{\epsilon_t}
~\partial_x \partial_y \Gr^{z \bar z}_{bou,~reg~U(t_a)}(x; y; \{x=0, x=\infty\})
\right]
}\Big|_{x = y= x_f}
\end{align}
in order to undo the OPE and get the desired result as in eq. (\ref{SDS-x=0}).

We shall now translate the previous operator (\ref{op-T})
into an abstract operator we can insert in the path integral 
at an arbitrary point $x_t$,
therefore we move it from $x_t=0$ to a generic $x_t$ 
%the previous operator
and we consider a generating vertex as 
%%%%%%%%%%%
\COMMENTOO{
1)This is the expression for $x>0$ while for $x<0$ the dipole
string would feel a  different magnetic field $B_{(t-1)}$.
\\
2)Explain why $x$ and not $x-x_t$ in $Z(x,x)$
\\
3) No ${(t)}$ because these are classical fields.
}
%%%%%%%%%%%
\begin{align}
\label{vertex-T}
\cT(d_{(t)},x_t)
=&
\lim_{x\rightarrow x_t^+}
\cN_\cT(d_{(t)}, x, x_t)
~
e^{
\bar d_{(t) 0 }  ~\langle Z(x, \bx)\rangle 
+
d_{(t) 0}  ~\langle  \bar   Z(x, \bx)\rangle 
}
\nonumber\\
&
e^{
\sum_{n=1}^\infty
\bar d_{(t) n }  
~\partial^{n-1}_x 
\langle (x-x_t)^{1-\epsilon_t}  \partial_x Z(x, \bx) \rangle 
+
d_{(t) n }  
~\partial^{n-1}_x 
\langle (x-x_t)^{\epsilon_t}  \partial_x \bar  Z(x, \bx)\rangle 
}
\end{align}
where 
$x\rightarrow x_t$ has to be understood as taking the limit after
the path integral has been computed.
We have defined the averaged fields such as
\begin{equation}
\langle (x-x_t)^{1-\epsilon_t}  \partial_x Z(x, \bx) \rangle
= \int_{\partial H} d y~ \delta_{reg}(x-y) (y-x_t)^{1-\epsilon_t}  
\partial_y Z(y, \by)
\end{equation}
because we want a well defined regulated expression after performing
the path integral
and introduced the normalization factor
\begin{align}
\cN_\cT(d_{(t)}, x, x_t)
=&
e^{
-
\oh \bar d_{(t) 0 }^2 \langle\langle G^{z z}_{N=2~bou}(x; x) \rangle\rangle
-
\oh  d_{(t) 0 }^2 \langle\langle G^{\bar z \bar z}_{N=2~bou}(x; x) \rangle\rangle
-
 \bar d_{(t) 0 } d_{(t) 0 }  
\langle\langle 
%\frac{ 
G^{z \bar z}_{N=2~bou}(x; x)  %+ G^{z \bar z}_{N=2~bou}(x; x) }{2}
\rangle\rangle
}
\nonumber\\
&
e^{
-\oh
\sum_{n=1}^\infty
\bar d_{(t) 0 } d_{(t) n}
\partial^{n-1}_x 
\langle\langle (x-x_t)^{\epsilon_t} \partial_x  G^{\bar z z}_{N=2~bou}(x; y) \rangle\rangle
}
\nonumber\\
&
e^{
-\oh
\sum_{n=1}^\infty
d_{(t) 0 } \bar d_{(t) n}
\partial^{n-1}_x 
\langle\langle (x-x_t)^{1-\epsilon_t} \partial_x  G^{z \bar z}_{N=2~bou}(x; y) \rangle\rangle
}
\nonumber\\
&
e^{
-\oh
\sum_{n,l=1}^\infty
d_{(t) l } \bar d_{(t) n}
\partial^{l-1}_y
\partial^{n-1}_x
\langle\langle 
(y-x_t)^{\epsilon_t} (x-x_t)^{1-\epsilon_t} \partial_x \partial_y  G^{z \bar  z}_{N=2~bou}(x; y)
\rangle\rangle
}
\Big|_{y=x}
\end{align}
where the doubly regularized Green functions are defined such as
\begin{align}
\langle\langle G^{z z}_{N=2~bou}(x; x) \rangle\rangle
&=
\int d y_1 \int d y_2 ~ \delta_{reg}(x-y_1) ~ \delta_{reg}(x-y_2)
\nonumber\\
&\hspace{7em}
G^{z z}_{bou}(y_1; y_2; \{x_1=x_t, x_2=\infty\})
\nonumber\\
\langle\langle 
(y-x_t)^{\epsilon_t} (x-x_t)^{1-\epsilon_t} \partial_x \partial_y  G^{z \bar  z}_{N=2~bou}(x; y)
\rangle\rangle
&=
\int d y_1 \int d y_2 
~\delta_{reg}(x-y_1) ~ \delta_{reg}(y-y_2)
\nonumber\\
&
%\hspace{7em}
(y_2-x_t)^{\epsilon_t}~ (y_1-x_t)^{1-\epsilon_t} 
\partial_1 \partial_2  G^{z \bar  z}_{bou}(y_1; y_2; \{x_1=x_t, x_2=\infty\}
)
\end{align} 
%%%%%%%%%
\COMMENTOO{
Could it be possible to do all the computations in operatorial
formalism?
Not really because I have no control over the Green function.
}
%%%%%%%%%%%
Since the previous expression is in nuce the same as for the untwisted
matter, we can immediately deduce from (\ref{PI_untwisted_1z})
the result of inserting and integrating over the $X$ 
to be eq. (\ref{Final-Reggeon-N+M-using-cd}).

In analogy with what done for the untwisted states we can realize the
algebra
\begin{equation}
\left[ d_{(t) n}, 
\frac{ \stackrel{\leftarrow}{\partial} }{ \partial d_{(u) m}}
\right]
=
\left[ \bar d_{(t) n}, 
\frac{ \stackrel{\leftarrow}{\partial} }{ \partial \bar d_{(u) m}}
\right]
= \delta_{m,n}` \delta_{u,t}
\end{equation} 
with  operators acting on 
the twisted scalar (dicharged string) auxiliary Hilbert spaces $\cH_{t}$
as
\begin{align}
\label{d-realization}
1 
\rightarrow &
\langle  T_{\epsilon_t}, x^1_{(t) 0}=0 |
\nonumber\\
\bar d_{(t) n\,} 
\rightarrow \frac{i}{\sqrt{2\alpha'}  \cos \gamma_{t}}
\frac{\alpha_{(t) n-1+\epsilon_t}}{(n-1)! ~(n-1+\epsilon_t)}
,&~~
d_{(t) n\, } 
\rightarrow \frac{i}{\sqrt{2\alpha'}  \cos \gamma_{t}}
\frac{\bar \alpha_{(t) n-\epsilon_t}}{(n-1)!~(n-\epsilon_t)}
~~~ n> 0
\nonumber\\
\frac{ \stackrel{\leftarrow}{\partial} }{ \partial  \bar d_{(t) m}}
\rightarrow
-i \sqrt{2 \alpha'}\, (m-1)!\, \cos \gamma_{t} \, \alpha_{(t) m-1+\epsilon_t}^\dagger
,&~~
\frac{ \stackrel{\leftarrow}{\partial} }{ \partial  d_{(t) m }}
\rightarrow
-i \sqrt{2 \alpha'}\, (m-1)!\, \cos \gamma_{t}\, 
\bar \alpha_{(t) m-\epsilon_t}^\dagger
~~ m>0
\end{align}
which gives eq. (\ref{Final-Reggeon-N+M}) 
when substituted into eq. (\ref{Final-Reggeon-N+M-using-cd}).

%%%%%%%%%%%%%%%%%%%%%%%%%%%%%%%%%%%%%%%%%%%%%%%%%%%%%%%%%%%%%%%%%%%%%%
%%%%%%%%%%%%%%%%%%%%%%%%%%%%%%%%%%%%%%%%%%%%%%%%%%%%%%%%%%%%%%%%%%%%%%
%%%%%%%%%%%%%%%%%%%%%%%%%%%%%%%%%%%%%%%%%%%%%%%%%%%%%%%%%%%%%%%%%%%%%%
%%%%%%%%%%%%%%%%%%%%%%%%%%%%%%%%%%%%%%%%%%%%%%%%%%%%%%%%%%%%%%%%%%%%%%

\noindent {\large {\bf Acknowledgments}}
\vskip 0.2cm
\noindent 
We would like to thank P. Di Vecchia and
F. Pezzella for discussions. The author thanks
the Nordita for hospitality during different stages of
this work.

%%%%%%%%%%%%%%%%%%%%%%%%%%%%%%%%%%%%%%%%%%%%%%%%%%%%%%%%%%%%%%%%%%%%%%
%%%%%%%%%%%%%%%%%%%%%%%%%%%%%%%%%%%%%%%%%%%%%%%%%%%%%%%%%%%%%%%%%%%%%%
%%%%%%%%%%%%%%%%%%%%%%%%%%%%%%%%%%%%%%%%%%%%%%%%%%%%%%%%%%%%%%%%%%%%%%
%%%%%%%%%%%%%%%%%%%%%%%%%%%%%%%%%%%%%%%%%%%%%%%%%%%%%%%%%%%%%%%%%%%%%%
\appendix

\section{Check of the $N=2$ amplitudes}
\label{app:N=2-check}
We would now check that the operatorial amplitudes with $N=2$ and the
path integral approach give the same result, phases included.
Let us consider the tachyonic amplitude
\begin{align}
\langle 
\sigma_{-\epsilon, \lambda}(x_\infty, x_\infty)
~\sigma_{\epsilon, \kappa}(x_0, x_0)
~V_T(x_1; k_{(1)})\dots ~V_T(x_M; k_{(M)})
\rangle
\end{align}
with $x_{t=1}=x_0$, $x_{t=2}=x_\infty$
and
$\gamma_0$ and $\gamma_1$ arbitrary but $\gamma_2=\gamma_0$ so that 
$\pi \epsilon_1=-\pi \epsilon_2$.  
Using the results from (\cite{Pesando:2011yd}) we can compute it
in the limit $x_0\rightarrow 0$ and $x_\infty \rightarrow \infty$,
for $x_1> \dots x_M>0$ and when multiplied by the appropriate power of
$x_\infty $ as
\begin{align}
&\langle T_{\epsilon}, -\lambda|
e^{ -\oh R^2(\epsilon) ~\Delta(k_{(1)})} x_1^{-\Delta(k_{(1)})} 
e^{i ( \bar k_{(1)} z_0 + k_{(1)} \bar z_0 )}
e^{i\cos\gamma_1 [  \bar k_{(1)} Z_{n z m}(x_1, x_1) + k_{(1)} \bar Z_{n
  z m}(x_1, x_1)] }
\dots
|T_{\epsilon}, \kappa\rangle
\nonumber\\
=
&
\delta(\kappa+\kappa+ \sum_a  k_{(a) 2} )
\nonumber\\
&
\prod_a 
e^{\oh \pi\alpha' \frac{1}{\tan\gamma_1 - \tan\gamma_0} ( k_{(a)}^2 - \bar k_{(a)}^2 ) }
\nonumber\\
&
\prod_a 
\Big[
e^{ -\oh R^2(\epsilon) ~\Delta(k_{(a)})}  
x_a^{-\Delta(k_{(a)})} 
\Big]
\nonumber\\
&
\prod_a
e^{- \pi\alpha' \frac{1}{\tan\gamma_1 - \tan\gamma_0}  \frac{\kappa}{\sqrt 2}( k_{(a)} - \bar k_{(a)} ) }
\nonumber\\
&
\prod_{a<b} 
\Big[
e^{ \pi\alpha' \frac{1}{\tan\gamma_1 - \tan\gamma_0} ( k_{(a)} - \bar k_{(a)} ) ( k_{(b)} + \bar k_{(b)} ) }
e^{ \alpha' \cos^2\gamma_1 
( k_{(a)} \bar k_{(b)} \,g_{1-\epsilon}\left(\frac{x_b}{x_a}\right) 
+\bar k_{(a)} k_{(b)}   \,g_{\epsilon}\left(\frac{x_b}{x_a}\right) 
) 
}
\Big]
\end{align}
where we have used the commutation relations
(\ref{comm-rel-dicharged}),
$ R^2(\epsilon)= 
\lim_{u\rightarrow 1^-}
[\,g_{\epsilon}(u) +\,g_{1-\epsilon}(u)
-2\log(1-u)]
=
-\left(
\psi(\epsilon)+\psi(1-\epsilon)-2\psi(1)
\right)
$
and
$~\Delta(k_{(a)})= 2 \alpha' \cos^2\gamma ~k_{(a)} \bar   k_{(a)}$
is the conformal dimension of the tachyonic vertex.

We can now compare with the general expression
(\ref{Final-Reggeon-N+M-using-cd})
with the identifications
$c_{(a)0} \rightarrow i~k_{(a)}$, 
$\bar c_{(a)0} \rightarrow i~\bar k_{(a)}$,
$d_{(1)0} \rightarrow i~\frac{\kappa}{\sqrt{2}}$,
$\bar d_{(1)0} \rightarrow i~\frac{\kappa}{\sqrt{2}}$,
$d_{(2)0} \rightarrow i~\frac{\lambda}{\sqrt{2}}$
and
$\bar d_{(2)0} \rightarrow i~\frac{\lambda}{\sqrt{2}}$.
We can also compare with the expression (\ref{Final-Reggeon-N+M}) upon
the product with the state...

in order to understand where the different terms come from the
path integral point of view.
In matching these terms is important to be careful in rewriting all
the Green functions $G(x;y)$ in such a way that $x>y$ by using the
symmetry (\ref{Green}) since this is the natural way they appear from
the operatorial formalism. 
We recognize that 
\begin{itemize}
\item
the factor 
$\exp\{\oh \pi\alpha' \frac{1}{\tan\gamma_1 - \tan\gamma_0} ( k_{(a)}^2 - \bar k_{(a)}^2 ) \}$
come from the $\Gr^{z z}_{bou~reg~U (t_a)}$ and $\Gr^{\bar z \bar z}_{bou~reg~U (t_a)}$
terms.
\item
The factors
$\prod_a \Big[ e^{ -\oh R^2(\epsilon) ~\Delta(k_{(a)})}  x_a^{-\Delta(k_{(a)})} 
\Big] $ come from the $\Gr^{z \bar z}_{bou~reg~U (t_a)}$ terms. 
In particular the result follows from the following steps
\begin{align}
\Gr^{z \bar z}_{bou~reg~U (t_a)}(x_a^+,x_a)
=&
\frac{\pi \alpha'}{\tan \gamma_1 -\tan \gamma_0} 
- \pi \alpha' \sin \gamma_1 \cos \gamma_1
+ 2\alpha' \cos^2 \gamma_1 \ln|x_a|
\nonumber\\
&
- 2\alpha' \cos^2 \gamma_1
\left(
\,g_\epsilon\left( \frac{x_a}{x_a^+} \right)
-\log\left(1- \frac{x_a}{x_a^+} \right)
\right)
\nonumber\\
&
=
\frac{\pi \alpha'}{\tan( \gamma_1 - \gamma_0)}  \cos^2 \gamma_1
+ 2\alpha' \cos^2 \gamma_1 \ln|x_a|
\nonumber\\
&
~~
- 2\alpha' \cos^2 \gamma_1
\left(
\psi(1-\epsilon)-\psi(1)
\right)
\nonumber\\
&=
2\alpha' \cos^2 \gamma_1 \ln|x_a|
- \alpha' \cos^2 \gamma_1
\left(
\psi(\epsilon)+\psi(1-\epsilon)-2\psi(1)
\right)
\end{align}
where we have used the expression for the Green function given in
eq. (\ref{twisted-bou-green}) since we have chosen $x=x_a^+$
and in the last line we have used the digamma property
$\psi(1-\epsilon)=\psi(\epsilon)+ \pi \cot (\pi \epsilon) $.
It is also worth stressing that the result is independent on setting
$x_a^+$
in the first argument
since the function $\Gr^{z \bar z}_{bou~reg~U (t_a)}$ is continued analytically at
$x=y=x_a$ in such a way that $G^{i j}(x;y)=G^{j i}(y; x)$
so that would we have chosen the first argument to be
$x_a^-$ we would  have got the same result computing 
$\Gr^{\bar z z}_{bou~reg~U (t_a)}(x_a,x_a^-)$.

\item
The terms $e^{- \pi\alpha' \frac{1}{\tan\gamma_1 - \tan\gamma_0}  \frac{\kappa}{\sqrt 2}(
  k_{(a)} - \bar k_{(a)} ) }$ arise from the $\prod_{t,a}$ terms.
While rewriting the Green functions $G(x;y)$ in such a way that $x>y$
we see that the terms with $t=2$ (at $x_\infty$) cancel 
while those from $t=1$ (at $x_0$) do not and reproduce the operatorial result.

\item
The terms $\prod_{a<b}$ come trivially from the corresponding ones in
eq. (\ref{Final-Reggeon-N+M-using-cd}).

\item
The terms $\prod_{t}$ in (\ref{Final-Reggeon-N+M-using-cd}) give a
trivial result in a non trivial way. It is immediate to find that
$G^{z z}_{bou,~reg~(t)}=G^{\bar z \bar z}_{bou,~reg~(t)}=0$. On the
other side we get for $0< \frac{y-x_0}{x-x_0}<1$ and 
$0<\omega= \frac{y-x_0}{x-x_0} \frac{x-x_\infty}{y-x_\infty}<1$
\begin{align}
G^{z \bar z}_{bou,~reg~(t)}(x; y; \{x_0,x_\infty\})
&=
-2\alpha' \cos^2\gamma \left[ \,g_\epsilon(\omega) -
\,g_\epsilon(\frac{y-x_0}{x-x_0}) \right]
\end{align}
Now we can write $x=x_0+\alpha$, $y=x_0+\alpha(1-\delta)$ with
$\alpha>0$ and $0<\delta<1$ and expand in $\delta$ to get
\begin{align}
G^{z \bar z}_{bou,~reg~(t)}(x; y; \{x_0,x_\infty\})
&=
-2\alpha' \cos^2\gamma \left[ 
-\log\left( 1+ \frac{\alpha}{x_\infty-x_0} \right)
+\frac{(\epsilon-1) \alpha }{x_\infty-x_0 +\alpha} \delta
+O(\delta^2) 
 \right]
\end{align}
which vanishes when $\alpha=\delta=0$. In a similar way can be
treated the case when $x,y \rightarrow x_\infty$.
\item
The terms $\prod_{t<u}$ in (\ref{Final-Reggeon-N+M-using-cd}) give
also a trivial result in a not completely trivial fashion.
We have to evaluate the Green functions for $\omega=
\frac{y-x_0}{x-x_0} \frac{x-x_\infty}{y-x_\infty}$  when
$x=x_0$ and $y=x_\infty$ so $\omega=0$ and we are left with only the
constant terms,as we expect from the general asymptotic
(\ref{Green-asymptotic})  hence
$\prod_{t<u} = \exp\{  \frac{\pi\alpha'}{\tan\gamma_1 - \tan\gamma_0}\
[ -d_{(1) 0}  d_{(2) 0} + \bar d_{(1) 0} \bar d_{(2) 0} 
-d_{(1) 0} \bar d_{(2) 0} +\bar d_{(1) 0}  d_{(2) 0} ]
\}$
which vanishes when evaluated with the previously stated substitutions
for which $d_{(t) 0}=\bar  d_{(t) 0} $.

\end{itemize}

\section{Behavior of the Green function when $x,y \rightarrow x_t$}
\label{Green-x-y-xt}
In this section we follow and adapt the computation done in
(\cite{Abel:2003yx}).
We start  considering the derivative of Green function of the left
moving part defined as
\begin{equation}
\label{Green-LL}
\partial_z \partial_w
\Gr^{z \bar z}_{L L}(z; w ; \{x_t\}_{t=1\dots N})
=
\frac{
\langle \partial_z Z_L(z)  \partial_w \bar Z_L (w)
\sigma_{\epsilon_1, \kappa_1}(x_1, \bar x_1)
\dots \sigma_{\epsilon_N, \kappa_N}(x_N, \bar x_N)\rangle_{disk}
}{
\langle \sigma_{\epsilon_1, \kappa_1}(x_1, \bar x_1)\dots
\sigma_{\epsilon_N, \kappa_N}(x_N, \bar x_N)\rangle_{disk}
}
\end{equation}
which has asymptotics
\begin{align}
-\frac{1}{2\alpha'}
\partial_z \partial_w
\Gr^{z \bar z}_{L L}(z; w ; \{x_t\}_{t=1\dots N})
&
\sim_{z\rightarrow w}
\frac{1}{(z-w)^2}+ O(1)
\nonumber\\
&
\sim_{z\rightarrow x_t}
(z-x_t)^{\epsilon_t-1}
\nonumber\\
&
\sim_{w\rightarrow x_t}
(w-x_t)^{-\epsilon_t}
\end{align}
then we can write
\begin{align}
-\frac{1}{2\alpha'}
\partial_z \partial_w
\Gr^{z \bar z}_{L L}(z; w ; \{x_t\}_{t=1\dots N})
=
&
\prod_u \left( \frac{w-x_u}{z-x_u} \right)^{1-\epsilon_u}
\Big[
\frac{1}{(z-w)^2}
\sum_{u<v} a_{u v} \frac{ (z-x_u) (z-x_v) }{ (w-x_u) (w-x_v)}
\nonumber\\
&
+
\sum_{u_1<u_2<u_3<u_4} \frac{b_{u_1 u_2 u_3 u_4} }
{ (w-x_{u_1}) (w-x_{u_2}) (w-x_{u_3}) (w-x_{u_4})  }
\Big]
\end{align}
Now we can study the behavior $x,y \rightarrow x_1$ by setting
\begin{equation}
z=x_1+\alpha,~~~~
w=x_1+\alpha(1-\delta)
\end{equation}
and letting $\alpha,\delta\rightarrow 0^+$.
A simple computation gives
\begin{align}
-\frac{1}{2\alpha'}
\partial_z \partial_w
\Gr^{z \bar z}_{L L}(z; w ; \{x_t\}_{t=1\dots N})
\sim
&
\frac{\sum_{u<v } a_{u v}}{ \delta^2}
+\frac{1}{\alpha} \frac{1}{1-\delta} 
\sum_{1<u_2<u_3<u_4} \frac{b_{1 u_2 u_3 u_4} }
{ (x_1-x_{u_2}) (x_1-x_{u_3}) (x_1-x_{u_4})  }
+O(1)
\end{align}
from which we deduce the constraint 
\begin{equation}
\sum_{u<v } a_{u v}=1.
\end{equation}
Then the regularized Green function which is obtained by subtracting
the corresponding Green function with only two twist, one of which in
$x_1$ and the other at $\infty$ is of the form
\begin{align}
\partial_z \partial_w
\Gr^{z \bar z}_{L L, ~ reg}(z; w ; \{x_t\}_{t=1\dots N})
\sim
&
\frac{B_1(x_u)}{\alpha} \frac{1}{1-\delta} 
+O(1)
\end{align}
hence the terms in $\prod_t$  are well defined since
\begin{align}
(z-x_1)^{1-\epsilon_1}~
(w-x_1)^{\epsilon_1}~
\partial_z \partial_w
\Gr^{z \bar z}_{L L, ~ reg}(z; w ; \{x_t\}_{t=1\dots N})
\sim
&
\alpha~ (1-\delta)^{\epsilon_1}
\left[
\frac{B_1(x_u)}{\alpha} \frac{1}{1-\delta} 
+O(1)
\right]
.
\end{align}
%We can then integrate this last expression to get the leading behavoir
%in the case of interest as

%%%%%%%%%%%%%%%%%%%%%%%%%%%%%%%%%%%%%%%%%%%%%%%%%%%%%%%%%%%%%%%%%%%%%%
%%%%%%%%%%%%%%%%%%%%%%%%%%%%%%%%%%%%%%%%%%%%%%%%%%%%%%%%%%%%%%%%%%%%%%
%%%%%%%%%%%%%%%%%%%%%%%%%%%%%%%%%%%%%%%%%%%%%%%%%%%%%%%%%%%%%%%%%%%%%%
%%%%%%%%%%%%%%%%%%%%%%%%%%%%%%%%%%%%%%%%%%%%%%%%%%%%%%%%%%%%%%%%%%%%%%

\end{document}